\documentclass[12pt]{article}
\usepackage{amssymb,amsfonts}
\usepackage{epsf,epsfig}
\begin{document}
\baselineskip 18pt

\title{Bethe ansatz for two-magnon scattering states in 2D and 3D Heisenberg-Ising ferromagnets}
\author{P.~N.~Bibikov}
\date{\it Russian State Hydrometeorological University, Saint-Petersburg, Russia}

%\affiliation{Saint-Petersburg State University}
\maketitle

\vskip5mm

\begin{abstract}
Two different versions of Bethe ansatz are suggested for evaluation of scattering two-magnon states in 2D and 3D Heisenberg-Ising ferromagnets on square and simple cubic lattices. It is shown that the two-magnon sector is subdivided on two subsectors related to non-interacting and scattering magnons. The former subsector possess an integrable regular dynamics and may be described by a natural modification of the usual Bethe Ansatz. The latter one is characterized by a non-integrable chaotic dynamics and may be treated only within discrete degenerative version of Bethe Ansatz previously suggested by the author. Some of these results are generalized for multi-magnon states of the Heisenberg-Ising ferromagnet on a $D$ dimensional hyper cubic lattice.
\end{abstract}

\maketitle

\section{Introduction}
It is known that the Bethe ansatz (BA) in its traditional version works well only for {\it integrable} 1D models giving exact representations for complete sets of many-particle wave functions \cite{1,2,3,4}. An extension of this approach to {\it non-integrable} systems may be suggested by two different ways.

First of all one may try to find even in a non-integrable case a limited number of states whose wave functions have forms inherent in BA. This
program was realized for three-magnon sectors of $S=1$ spin chains \cite{5} rung-dimerized spin ladders \cite{6} and for many other {\it partially integrable} quantum mechanical models \cite{7}.

The recently suggested alternative approach \cite{8} based on the {\it degenerative discrete-diffractive Bethe ansatz} (DDD BA) is elaborated especially for {\it non-integrable} models.
In particular it was used for evaluation of the overcomplete set of three-magnon states in general ({\it non-integrable}) isotropic $S=1$ ferromagnetic chain \cite{8}.

In the present paper within the two mentioned approaches we study the two-magnon problem in 2D and 3D Heisenberg-Ising (XXZ) ferromagnets on infinite and finite-periodic square and simple cubic lattices. It is well known that both these models are non-integrable \cite{9}. Nevertheless we show that their two-magnon sectors may be naturally subdivided on two orthogonal subsectors which we call antisymmetric and symmetric.
The former one may be completely described within the partially integrability approach either for a finite-periodic or an infinite lattices.  Namely both in 2D and 3D we have obtained complete orthogonal systems of states and showed that they correspond to {\it non-interacting} magnons. This result indicates that these systems are partially integrable in the sense of Ref. 7. In its turn for the remained symmetric sector we have applied the DDD version of BA which is however effective only for infinite lattices. With its use we have obtained an overcomplete system of {\it interacting} scattering states. At the same time any description of resonant and bound states \cite{10,11,12} (which also lie in the symmetric sector) in the framework of DDD BA is unclear for the author.

The paper is organized as follows. In Sect. 2 we suggest the 2D XXZ model on an infinite square lattice rewriting all one-magnon states which are the usual Bloch waves \cite{9} and deriving a complete orthogonal basis in the antisymmetric two-magnon subsector. The corresponding wave functions are linear combinations of {\it four} (not {\it two} as in 1D!) exponents, which is the correct generalization of the 1D BA to the 2D case. We explain why the antisymmetric subsector contains just the non-interacting states. In Sect. 3 we treat the symmetric subsector within the DDD BA and obtain an overcomplete system of {\it interacting} two-magnon scattering states. In Sect. 4 we generalize the results of sections 2 and 3 to the 3D model on a simple cubic lattice.  In Sect. 5 we give some general remarks on the DDD BA and explain when it is effective and when is not. The Sections 6 and 7 are devoted to generalization of the results presented in Sections 2-4. In Sect. 6 we study the two-magnon problem on a $D$-dimensional hyper cubic lattice. In Sect. 7 some of these results are generalized for $(Q\geq3)$-magnon $D$-dimensional states. In Sect. 8 we compare our approach with two traditional ones \cite{9,13,14}. Finitely in the Sect. 9 we give a summary.

\section{Antisymmetric two-magnon Bethe states on 2D square infinite lattice}

The 2D square lattice Heisenberg-Ising Hamiltonian has the form
\begin{eqnarray}
&&\hat H=-\sum_{m,n=-\infty}^{\infty}\Big[\frac{J_1}{2}\Big({\bf S}^-_{m-1,n}+{\bf S}^-_{m+1,n}\Big){\bf S}^+_{m,n}+
\frac{J_2}{2}\Big({\bf S}^-_{m,n-1}+{\bf S}^-_{m,n+1}\Big){\bf S}^+_{m,n}\nonumber\\
&&+J_{z,1}\Big({\bf S}^z_{m,n}{\bf S}^z_{m+1,n}-\frac{1}{4}\Big)+J_{z,2}\Big({\bf S}^z_{m,n}{\bf S}^z_{m,n+1}-\frac{1}{4}\Big)+\gamma h\Big({\bf S}_{m,n}^z-\frac{1}{2}\Big)\Big],
\end{eqnarray}
where ${\bf S}_{m,n}^z$ and ${\bf S}_{m,n}^{\pm}={\bf S}_{m,n}^x\pm i{\bf S}_{m,n}^y$ are the usual spin-1/2 operators.
At $J_{z,1},J_{z,1}>0$ or at rather strong $h>0$ Hamiltonian (1) has the spin-polarized ground state
\begin{equation}
|\emptyset\rangle=\prod_{m,n=-\infty}^{\infty}\otimes|\uparrow\rangle_{m,n},
\end{equation}
where $|\uparrow\rangle_{m,n}$ and $|\downarrow\rangle_{m,n}$ are spin up and down polarized states related to $(m,n)$-th site.

The corresponding magnons are Bloch waves \cite{9}
\begin{equation}
|k,p\rangle=\sum_{m,n}{\rm e}^{i(km+pn)}{\bf S}^-_{m,n}|\emptyset\rangle,
\end{equation}
with energies
\begin{equation}
E(k,p,h)=E_1(k)+E_2(p)+\gamma h,
\end{equation}
where
\begin{equation}
E_1(k)=J_{z,1}-J_1\cos{k},\qquad E_2(p)=J_{z,2}-J_2\cos{p}.
\end{equation}

A two-magnon state should have the form
\begin{equation}
|2\rangle=\sum_{m_1,m_2,n_1,n_2}a_{m_1,m_2,n_1,n_2}{\bf S}^-_{m_1,n_1}{\bf S}^-_{m_2,n_2}|\emptyset\rangle,
\end{equation}
where without loss of generality one may postulate the following symmetry condition
\begin{equation}
a_{m_2,m_1,n_2,n_1}=a_{m_1,m_2,n_1,n_2},
\end{equation}
or equivalently
\begin{equation}
({\cal P}a)_{m_1,m_2,n_1,n_2}=\sum_{\tilde m_1,\tilde m_2,\tilde n_1,\tilde n_2}
{\cal P}_{m_1,m_2,n_1,n_2,\tilde m_1,\tilde m_2,\tilde n_1,\tilde n_2}a_{\tilde m_1,\tilde m_2,\tilde n_1,\tilde n_2}
=a_{m_1,m_2,n_1,n_2},
\end{equation}
where
\begin{equation}
{\cal P}_{m_1,m_2,n_1,n_2,\tilde m_1,\tilde m_2,\tilde n_1,\tilde n_2}=\frac{1}{2}
\Big(\delta_{m_1\tilde m_1}\delta_{m_2\tilde m_2}\delta_{n_1\tilde n_1}\delta_{n_2\tilde n_2}+\delta_{m_1\tilde m_2}\delta_{m_2\tilde m_1}
\delta_{n_1\tilde n_2}\delta_{n_2\tilde n_1}\Big),
\end{equation}
is an idempotent (projection operator)
\begin{equation}
{\cal P}^2={\cal P}.
\end{equation}

Since for $S=1/2$ one has $({\bf S}_{m,n}^-)^2=0$, the region
\begin{equation}
m_2=m_1,\qquad n_2=n_1,
\end{equation}
is unphysical and the values $a_{m,m,n,n}$ may be taken arbitrary. Hence we suggest the following scalar product in the Hilbert space of two-magnon states (6)
\begin{equation}
\langle2|\tilde2\rangle=\sum_{m_1,m_2,n_1,n_2=-\infty}^{\infty}
(1-\delta_{m_1m_2}\delta_{n_1n_2})\bar a_{m_1,m_2,n_1,n_2}
\tilde a_{m_1,m_2,n_1,n_2}.
\end{equation}

The corresponding ${\rm Schr\ddot odinger}$ equation splits on three subsystems
\begin{eqnarray}
&&2(J_{z,1}+J_{z,2}+\gamma h)a_{m_1,m_2,n_1,n_2}-\sum_{l=\pm1}\Big[\frac{J_1}{2}\Big(a_{m_1+l,m_2,n_1,n_2}+
a_{m_1,m_2+l,n_1,n_2}\Big)\nonumber\\
&&+\frac{J_2}{2}\Big(a_{m_1,m_2,n_1+l,n_2}+a_{m_1,m_2,n_1,n_2+l}\Big)\Big]
=Ea_{m_1,m_2,n_1,n_2},
\nonumber\\
&&(m_2-m_1)^2+(n_2-n_1)^2>1,\\
&&(J_{z,1}+2J_{z,2}+2\gamma h)a_{m,m+1,n,n}-\frac{J_1}{2}\Big(a_{m-1,m+1,n,n}+a_{m,m+2,n,n}\Big)\nonumber\\
&&-\frac{J_2}{2}\sum_{l=\pm1}\Big(a_{m,m+1,n+l,n}+a_{m,m+1,n,n+l}\Big)=Ea_{m,m+1,n,n},\\
&&(2J_{z,1}+J_{z,2}+2\gamma h)a_{m,m,n,n+1}-\frac{J_1}{2}\sum_{l=\pm1}\Big(a_{m+l,m,n,n+1}+a_{m,m+l,n,n+1}\Big)
\nonumber\\
&&-\frac{J_2}{2}\Big(a_{m,m,n-1,n+1}+a_{m,m,n,n+2}\Big)=Ea_{m,m,n,n+1}.
\end{eqnarray}

As usual \cite{1,2,3,4} one may expand the wave function $a_{m_1,n_1,m_2,n_2}$ into an unphysical region (11) and reduce both Eqs. (14) and (15) to the form (13) under the following Bethe conditions (their proof in $D$ dimensions is given in Eq. (173))
\begin{eqnarray}
&&J_1\Big(a_{m,m,n,n}+a_{m+1,m+1,n,n}\Big)=2J_{z,1}a_{m,m+1,n,n},\nonumber\\
&&J_2\Big(a_{m,m,n,n}+a_{m,m,n+1,n+1}\Big)=2J_{z,2}a_{m,m,n,n+1}.
\end{eqnarray}

Under the symmetry (7) the correct 2D generalization of the 1D Bethe two-magnon wave function should be
\begin{eqnarray}
&&a_{m_1,m_2,n_1,n_2}(k_1,k_2,p_1,p_2)=A\Big({\rm e}^{i(k_1m_1+k_2m_2+p_1n_1+p_2n_2)}+{\rm e}^{i(k_2m_1+k_1m_2+p_2n_1+p_1n_2)}\Big)
\nonumber\\
&&-\tilde A\Big({\rm e}^{i(k_2m_1+k_1m_2+p_1n_1+p_2n_2)}+{\rm e}^{i(k_1m_1+k_2m_2+p_2n_1+p_1n_2)}\Big).
\end{eqnarray}
The first two terms correspond to particles with the wave numbers $(k_1,p_1)$ and $(k_2,p_2)$ while the last two are related to the ones with the wave numbers $(k_1,p_2)$ and $(k_2,p_1)$. The wave function (17) solves Eq. (13) giving an energy
\begin{equation}
E(k_1,k_2,p_1,p_2,h)=E_1(k_1)+E_1(k_2)+E_2(p_1)+E_2(p_2)+2\gamma h.
\end{equation}
At the same time a substitution of (17) into (16) results in
\begin{eqnarray}
(A-\tilde A)X^{(1)}(k_1,k_2){\rm e}^{i[(k_1+k_2)m+(p_1+p_2)n]}=(A-\tilde A)X^{(2)}(p_1,p_2){\rm e}^{i[(k_1+k_2)m+(p_1+p_2)n]}=0,
\end{eqnarray}
where
\begin{equation}
X^{(l)}(w_1,w_2)=J_l\Big(1+{\rm e}^{i(w_1+w_2)}\Big)-J_{z,l}\Big({\rm e}^{iw_1}+{\rm e}^{iw_2}\Big).
\end{equation}

Eq. (19) has a single solution $\tilde A=A$. If we put
\begin{equation}
\tilde A=A=\frac{1}{2},
\end{equation}
then the corresponding wave function takes the form
\begin{equation}
a^{(asym)}_{m_1,m_2,n_1,n_2}(k_1,k_2,p_1,p_2)
=a^{(a)}_{m_1,m_2}(k_1,k_2)a^{(a)}_{n_1,n_2}(p_1,p_2),
\end{equation}
or in invariant notation
\begin{equation}
a^{(asym)}(k_1,k_2,p_1,p_2)
=a^{(a)}(k_1,k_2)\otimes a^{(a)}(p_1,p_2),
\end{equation}
where
\begin{equation}
a^{(a)}_{m_1,m_2}(k_1,k_2)=\frac{1}{\sqrt{2}}\Big({\rm e}^{i(k_1m_1+k_2m_2)}-{\rm e}^{i(k_2m_1+k_1m_2)}\Big),
\end{equation}
is the two-magnon wave function of the 1D $XX$ model \cite{15}. It is antisymmetric under permutation of the two indices $m_1$ and $m_2$
\begin{equation}
a^{(a)}_{m_1,m_2}(k_1,k_2)=-a^{(a)}_{m_2,m_1}(k_1,k_2),
\end{equation}
or equivalently
\begin{equation}
\sum_{\tilde m_1,\tilde m_2}P^{(asym)}_{m_1,m_2,\tilde m_1,\tilde m_2}
a^{(a)}_{\tilde m_1,\tilde m_2}(k_1,k_2)=a^{(a)}_{m_1,m_2}(k_1,k_2),
\end{equation}
where
\begin{equation}
P^{(asym)}_{m_1,m_2,\tilde m_1,\tilde m_2}=\frac{1}{2}\Big(\delta_{m_1\tilde m_1}\delta_{m_2\tilde m_2}-\delta_{m_1\tilde m_2}\delta_{m_2\tilde m_1}\Big),
\end{equation}
are the matrix elements of the 1D asymmetric projector. Eq. (26) may be also represented in the compact form
\begin{equation}
P^{(asym)}a^{(a)}(k_1,k_2)=a^{(a)}(k_1,k_2).
\end{equation}

According to (28) the wave function (23) satisfies the condition
\begin{equation}
{\cal P}^{(asym)}a^{(asym)}(k_1,k_2,p_1,p_2)=a^{(asym)}(k_1,k_2,p_1,p_2).
\end{equation}
where
\begin{equation}
{\cal P}^{(asym)}_{m_1,m_2,n_1n_2,\tilde m_1,\tilde m_2,\tilde n_1,\tilde n_2}=
P^{(asym)}_{m_1,m_2,\tilde m_1,\tilde m_2}P^{(asym)}_{n_1,n_2,\tilde n_1,\tilde n_2},
\end{equation}
or in an invariant form
\begin{equation}
{\cal P}^{(asym)}=P^{(asym)}\otimes P^{(asym)}.
\end{equation}

Since
\begin{equation}
a^{(a)}_{m_1,m_2}(k_2,k_1)=-a^{(a)}_{m_1,m_2}(k_1,k_2),
\end{equation}
one also has
\begin{equation}
a^{(asym)}_{m_1,m_2,n_1,n_2}(k_2,k_1,p_1,p_2)=a^{(asym)}_{m_1,m_2,n_1,n_2}(k_1,k_2,p_2,p_1)=
-a^{(asym)}_{m_1,m_2,n_1,n_2}(k_1,k_2,p_1,p_2).
\end{equation}
Hence we may everywhere put
\begin{equation}
0\leq k_1<k_2<2\pi,\qquad0\leq p_1<p_2<2\pi.
\end{equation}
It may be readily proved that under this condition
\begin{equation}
\sum_{m_1,m_2=-\infty}^{\infty}\bar a^{(a)}_{m_1,m_2}(k_1,k_2)a^{(a)}_{m_1,m_2}(\tilde k_1,\tilde k_2)=(2\pi)^2
\delta(\tilde k_1-k_1)\delta(\tilde k_2-k_2),
\end{equation}
and according to (12), (25) and (22)
\begin{eqnarray}
&&\langle\tilde k_1,\tilde k_2,\tilde p_1,\tilde p_2,asym|k_1,k_2,p_1,p_2,asym\rangle
=\sum_{m_1,m_2,n_1,n_2=-\infty}^{\infty}
\bar a^{(asym)}_{m_1,m_2,n_1,n_2}\tilde a^{(asym)}_{m_1,m_2,n_1,n_2}\nonumber\\
&&=(2\pi)^4\delta(\tilde k_1-k_1)\delta(\tilde k_2-k_2)\delta(\tilde p_1-p_1)\delta(\tilde p_2-p_2).
\end{eqnarray}

The wave functions (24) give a resolution of $P^{(asym)}$. In fact, according to (34)
\begin{eqnarray}
&&\frac{1}{(2\pi)^2}\int_0^{2\pi}dk_2\int_0^{k_2}dk_1
\bar a^{(a)}_{m_1,m_2}(k_1,k_2)a^{(a)}_{\tilde m_1,\tilde m_2}(k_1,k_2)
\nonumber\\
&&=\frac{1}{16\pi^2}\int_0^{2\pi}dk_2\int_0^{2\pi}dk_1\Big({\rm e}^{-i(k_1m_1+k_2m_2)}-
{\rm e}^{-i(k_2m_1+k_1m_2)}\Big)\Big({\rm e}^{i(k_1\tilde m_1+k_2\tilde m_2)}-
{\rm e}^{i(k_2\tilde m_1+k_1\tilde m_2)}\Big)\nonumber\\
&&=\frac{1}{2}\Big(\delta_{m_1\tilde m_1}\delta_{m_2\tilde m_2}-\delta_{m_1\tilde m_2}\delta_{m_2\tilde m_1}\Big)
=P^{(asym)}_{m_1,m_2,\tilde m_1,\tilde m_2}.
\end{eqnarray}
Form this formula and Eqs. (22), (30) follows that
\begin{eqnarray}
&&\frac{1}{(2\pi)^4}\int_0^{2\pi}dk_2\int_0^{k_2}dk_1\int_0^{2\pi}dp_2\int_0^{p_2}dp_1
\bar a^{(asym)}_{m_1,m_2,n_1,n_2}(k_1,k_2,p_1,p_2)a^{(asym)}_{\tilde m_1,\tilde m_2,\tilde n_1,\tilde n_2}(k_1,k_2,p_1,p_2)\nonumber\\
&&={\cal P}^{(asym)}_{m_1,m_2,n_1,n_2,\tilde m_1,\tilde m_2,\tilde n_1,\tilde n_2}.
\end{eqnarray}
This equation means that the states $|k_1,k_2,p_1,p_2,asym\rangle$ related to the wave functions (22) give the resolution of ${\cal P}^{(asym)}$. Namely
\begin{equation}
\frac{1}{(2\pi)^4}\int_0^{2\pi}dk_2\int_0^{k_2}dk_1\int_0^{2\pi}dp_2\int_0^{p_2}dp_1|k_1,k_2,p_1,p_2,asym\rangle\langle k_1,k_2,p_1,p_2,asym|={\cal P}^{(asym)}.
\end{equation}

Since the wave functions $a^{(asym)}_{m_1,m_2,n_1,n_2}(k_1,k_2,p_1,p_2)$ do not depend on the coupling parameters and according to the factorization formula (22) we may suggest that they describe a free motion of two non-interacting magnons which even do not feel each other. The latter statement also follows from the fact that $a^{(asym)}_{m_1,m_2,n_1,n_2}(k_1,k_2,p_1,p_2)$ turns to zero either at $m_1=m_2$ or at $n_1=n_2$ (see Fig. 1).

Postulating the periodic boundary conditions
\begin{equation}
a_{m_1+N_x,m_2,n_1,n_2}=a_{m_1,m_2,n_1+N_y,n_2}=a_{m_1,m_2,n_1,n_2},
\end{equation}
one may readily get the following quantization rules
\begin{equation}
{\rm e}^{ik_jN_x}={\rm e}^{ip_jN_y}=1,\qquad j=1,2,
\end{equation}
on the wave numbers of the antisymmetric scattering wave functions (22) related to periodic $N_x\times N_y$ lattice. Eqs. (41) define
\begin{equation}
N^{(asym)}=\frac{N_x(N_x-1)}{2}\cdot\frac{N_y(N_y-1)}{2}=\frac{N_xN_y(N_x-1)(N_y-1)}{4},
\end{equation}
states related to wave numbers
\begin{eqnarray}
&&k_1=\frac{2\pi j_1}{N_x},\quad k_2=\frac{2\pi j_2}{N_x},\qquad0\leq j_1<j_2<N_x-1,\quad j_1,j_2\in{\mathbb N},\nonumber\\
&&p_1=\frac{2\pi l_1}{N_y},\quad p_2=\frac{2\pi l_2}{N_y},\qquad0\leq l_1<l_2<N_y-1,\quad l_1,l_2\in{\mathbb N}.
\end{eqnarray}
Since $N(N-1)/2$ is the range of the operator $P^{(asym)}$ on a 1D chain with $N$ sites we see from Eq. (42) that
$N^{(asym)}$ is just the range of the operator ${\cal P}^{(asym)}$. In other words the system of Bethe states (6), (22) gives a complete and orthogonal basis of the antisymmetric subsector.

At $N_x,N_y\gg1$ the number of antisymmetric states $N^{(asym)}$ is about a half of the total number of two-magnon states. In fact in this limit
\begin{equation}
N^{(tot)}=\frac{N_xN_y(N_xN_y-1)}{2}\approx\frac{N_x^2N_y^2}{2},\qquad N^{(asym)}\approx\frac{N_x^2N_y^2}{4}\approx\frac{N^{(tot)}}{2}.
\end{equation}

\section{Symmetric scattering two-magnon Bethe states on 2D square infinite lattice}

The antisymmetric subsector related to the Bethe wave functions (22) is supplemented by its symmetric counterpart for which the relation (29) is changed on
\begin{equation}
{\cal P}^{(sym)}a^{(sym)}=a^{(sym)}.
\end{equation}
Here
\begin{equation}
{\cal P}^{(sym)}={\cal P}-{\cal P}^{(asym)},
\end{equation}
is the symmetric projector with the matrix elements
\begin{equation}
{\cal P}^{(sym)}_{m_1,m_2,n_1,n_2,\tilde m_1\tilde m_2,\tilde n_1,\tilde n_2}=\frac{1}{4}\Big(\delta_{m_1\tilde m_1}\delta_{m_2\tilde m_2}+\delta_{m_1\tilde m_2}\delta_{m_2\tilde m_1}\Big)
\Big(\delta_{n_1\tilde n_1}\delta_{n_2\tilde n_2}+\delta_{n_1\tilde n_2}\delta_{n_2\tilde n_1}\Big).
\end{equation}
In invariant notation
\begin{equation}
{\cal P}^{(sym)}=P^{(sym)}\otimes P^{(sym)},
\end{equation}
where
\begin{equation}
P^{(sym)}_{m_1,m_2,\tilde m_1,\tilde m_2}=\frac{1}{2}\Big(\delta_{m_1\tilde m_1}\delta_{m_2\tilde m_2}+\delta_{m_1\tilde m_2}\delta_{m_2\tilde m_1}\Big).
\end{equation}

In order to obtain an elementary symmetric wave function we should put $\tilde A=-A$ in (17). Taking by analogy with (21) $A=1/2$, $\tilde A=-1/2$ we get from (17)
\begin{equation}
a^{(sym)}_{m_1,m_2,n_1,n_2}(k_1,k_2,p_1,p_2)=a^{(s)}_{m_1,m_2}(k_1,k_2)
a^{(s)}_{n_1,n_2}(p_1,p_2),
\end{equation}
where the symmetric function
\begin{equation}
a^{(s)}_{m_1,m_2}(k_1,k_2)=\frac{1}{\sqrt{2}}\Big({\rm e}^{i(k_1m_1+k_2m_2)}+{\rm e}^{i(k_2m_1+k_1m_2)}\Big),
\end{equation}
satisfies the symmetric analog of Eq. (28)
\begin{equation}
P^{(sym)}a^{(s)}(k_1,k_2)=a^{(s)}(k_1,k_2).
\end{equation}

As it was shown in the previous section this wave function satisfies Eq. (13) giving the energy (18). However its substitution does not solves Eq. (16) but reduce it to a system
\begin{equation}
X^{(1)}(k_1,k_2)=X^{(2)}(p_1,p_2)=0.
\end{equation}

In order to get over this obstacle we suggest a discrete-diffractive wave function
\begin{eqnarray}
&&a^{(DDD,sym)}_{m_1,m_2,n_1,n_2}(\{k_1^{(1)},k_2^{(1)},p_1^{(1)},p_2^{(1)}\},\dots,\{k_1^{(M)},k_2^{(M)},p_1^{(M)},p_2^{(M)}\})\nonumber\\
&&=\sum_{j=1}^MB_ja^{(sym)}_{m_1,m_2,n_1,n_2}(k_1^{(j)},k_2^{(j)},p_1^{(j)},p_2^{(j)}),
\end{eqnarray}
where $B_j$ are some numbers. According to the energy and quasimomentum conservation laws for all $j=1,\dots,M$ there should be
\begin{eqnarray}
&&{\rm e}^{i(k_1^{(j)}+k_2^{(j)})}={\rm e}^{ik},\qquad
{\rm e}^{i(p_1^{(j)}+p_2^{(j)})}={\rm e}^{ip},\nonumber\\
&&E_1(k_1^{(j)})+E_1(k_2^{(j)})+E_2(p_1^{(j)})+E_2(p_2^{(j)})=E-2\gamma h.
\end{eqnarray}
The parameters $k$ and $p$ in (55) characterize the total wave numbers (components of the total quasimomentum along the two Cartesian axes).
We shall be interesting here only in the degenerative solutions for which
\begin{equation}
1<M<\infty.
\end{equation}

As a sum of exponents the wave function (54) satisfies Eq. (13). At the same time its substitution into (16) results in a system
\begin{equation}
\sum_{j=1}^MB_jX^{(1)}(k_1^{(j)},k_2^{(j)})=0,\qquad\sum_{j=1}^MB_jX^{(2)}(p_1^{(j)},p_2^{(j)})=0.
\end{equation}

From (20) and (55) follows that in the $XX$ case \cite{15}
\begin{equation}
J_{z,1}=J_{z,2}=0\Longrightarrow X^{(1)}(k_1^{(j)},k_2^{(j)})=J_1\Big(1+{\rm e}^{ik}\Big),\qquad X^{(2)}(p_1^{(j)},p_2^{(j)})=J_1\Big(1+{\rm e}^{ip}\Big),
\end{equation}
and both the equations in (57) turn into a single one
\begin{equation}
\sum_{j=1}^MB_j=0.
\end{equation}
The latter has the $M=2$ solution
\begin{equation}
B\equiv B_1=-B_2,
\end{equation}
related to the wave function
\begin{eqnarray}
&&a^{(DDD,sym,XX)}_{m_1,m_2,n_1,n_2}
(k_1,k_2,p_1,p_2,\tilde k_1,\tilde k_2,\tilde p_1,\tilde p_2)
=a^{(s)}_{m_1,m_2}(k_1,k_2)a^{(s)}_{n_1,n_2}(p_1,p_2)\nonumber\\
&&-a^{(s)}_{m_1,m_2}(\tilde k_1,\tilde k_2)a^{(s)}_{n_1,n_2}(\tilde p_1,\tilde p_2),
\end{eqnarray}
(where we have put $k^{(1)}_l\equiv k_l$, $k^{(2)}_l\equiv\tilde k_l$, $p^{(1)}_l\equiv p_l$, $p^{(2)}_l\equiv\tilde p_l$, $l=1,2$ and $B=1$).

In the general case $J_{z,1},J_{z,2}\neq0$, Eq. (57) has the $M=3$ solution
\begin{equation}
B_j=\sum_{l,m=1}^3\varepsilon_{jlm}X^{(1)}(k_1^{(l)},k_2^{(l)})X^{(2)}(p_1^{(m)},p_2^{(m)}),\qquad j=1,2,3.
\end{equation}

Since
\begin{equation}
a^{(s)}_{m_1,m_2}(k_2,k_1)=a^{(s)}_{m_1,m_2}(k_1,k_2),
\end{equation}
one has
\begin{eqnarray}
&&a^{(DDD,sym,XX)}_{m_1,m_2,n_1,n_2}(k_2,k_1,p_1,p_2,\tilde k_1,\tilde k_2,\tilde p_1,\tilde p_2)=
a^{(DDD,sym,XX)}_{m_1,m_2,n_1,n_2}(k_1,k_2,p_2,p_1,\tilde k_1,\tilde k_2,\tilde p_1,\tilde p_2)\nonumber\\
&&=a^{(DDD,sym,XX)}_{m_1,m_2,n_1,n_2}(k_1,k_2,p_1,p_2,\tilde k_2,\tilde k_1,\tilde p_1,\tilde p_2)
=a^{(DDD,sym,XX)}_{m_1,m_2,n_1,n_2}(k_1,k_2,p_1,p_2,\tilde k_1,\tilde k_2,\tilde p_2,\tilde p_1)
\nonumber\\
&&=a^{(DDD,sym,XX)}_{m_1,m_2,n_1,n_2}(k_1,k_2,p_1,p_2,\tilde k_1,\tilde k_2,\tilde p_1,\tilde p_2),
\end{eqnarray}
for the XX case (61) and
\begin{eqnarray}
&&a^{(DDD,sym)}_{m_1,m_2,n_1,n_2}(\dots,\{k_2^{(j)},k_1^{(j)},p_1^{(j)},p_2^{(j)}\},\dots)=
a^{(DDD,sym)}_{m_1,m_2,n_1,n_2}(\dots,\{k_1^{(j)},k_2^{(j)},p_2^{(j)},p_1^{(j)}\},\dots)\nonumber\\
&&=a^{(DDD,sym)}_{m_1,m_2,n_1,n_2}(\{k_1^{(1)},k_2^{(1)},p_1^{(1)},p_2^{(1)}\},
\{k_1^{(2)},k_2^{(2)},p_1^{(2)},p_2^{(2)}\},\{k_1^{(3)},k_2^{(3)},p_1^{(3)},p_2^{(3)}\}),
\end{eqnarray}
in the general case (54), (62).

Hence by analogy with (34) we may put
\begin{equation}
0\leq k_1<k_2<2\pi,\quad0\leq\tilde k_1<\tilde k_2<2\pi,\quad0\leq p_1<p_2<2\pi,\quad0\leq \tilde p_1<\tilde p_2<2\pi,
\end{equation}
in the XX case (61) and
\begin{equation}
0\leq k_1^{(j)}<k_2^{(j)}<2\pi,\qquad0\leq p_1^{(j)}<p_2^{(j)}<2\pi,\qquad j=1,2,3,
\end{equation}
in the general case (54), (62).

The following figures illustrate the differences between spin configurations available in antisymmetric and symmetric two-magnon subsectors.

\begin{picture}(200,300)
\put(0,160){\vector(1,0){70}}
\put(0,160){\vector(1,2){35}}
\put(20,170){\vector(0,1){10}}
\put(40,170){\vector(0,1){10}}
\put(60,170){\vector(0,1){10}}
\put(30,200){\vector(0,-1){10}}
\put(50,200){\vector(0,-1){10}}
\put(70,190){\vector(0,1){10}}
\put(40,210){\vector(0,1){10}}
\put(60,210){\vector(0,1){10}}
\put(80,210){\vector(0,1){10}}
\put(75,160){x}
\put(40,230){y}
\put(150,160){\vector(1,0){70}}
\put(150,160){\vector(1,2){35}}
\put(170,170){\vector(0,1){10}}
\put(190,180){\vector(0,-1){10}}
\put(210,170){\vector(0,1){10}}
\put(180,190){\vector(0,1){10}}
\put(200,200){\vector(0,-1){10}}
\put(220,190){\vector(0,1){10}}
\put(190,210){\vector(0,1){10}}
\put(210,210){\vector(0,1){10}}
\put(230,210){\vector(0,1){10}}
\put(225,160){x}
\put(190,230){y}
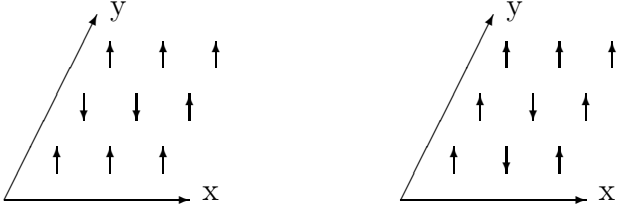
\put(0,140){Fig. 1 Such configurations of spins are possible only for the symmetric subsector.}
\put(0,40){\vector(1,0){70}}
\put(0,40){\vector(1,2){35}}
\put(20,50){\vector(0,1){10}}
\put(40,50){\vector(0,1){10}}
\put(60,50){\vector(0,1){10}}
\put(30,80){\vector(0,-1){10}}
\put(50,70){\vector(0,1){10}}
\put(70,80){\vector(0,-1){10}}
\put(40,90){\vector(0,1){10}}
\put(60,90){\vector(0,1){10}}
\put(80,90){\vector(0,1){10}}
\put(75,40){x}
\put(40,110){y}
\put(150,40){\vector(1,0){70}}
\put(150,40){\vector(1,2){35}}
\put(170,50){\vector(0,1){10}}
\put(190,60){\vector(0,-1){10}}
\put(210,50){\vector(0,1){10}}
\put(180,70){\vector(0,1){10}}
\put(200,70){\vector(0,1){10}}
\put(220,70){\vector(0,1){10}}
\put(190,90){\vector(0,1){10}}
\put(210,100){\vector(0,-1){10}}
\put(230,90){\vector(0,1){10}}
\put(225,40){x}
\put(190,110){y}
\put(300,40){\vector(1,0){70}}
\put(300,40){\vector(1,2){35}}
\put(320,50){\vector(0,1){10}}
\put(340,60){\vector(0,-1){10}}
\put(360,50){\vector(0,1){10}}
\put(330,80){\vector(0,-1){10}}
\put(350,70){\vector(0,1){10}}
\put(370,70){\vector(0,1){10}}
\put(340,90){\vector(0,1){10}}
\put(360,90){\vector(0,1){10}}
\put(380,90){\vector(0,1){10}}
\put(375,40){x}
\put(340,110){y}
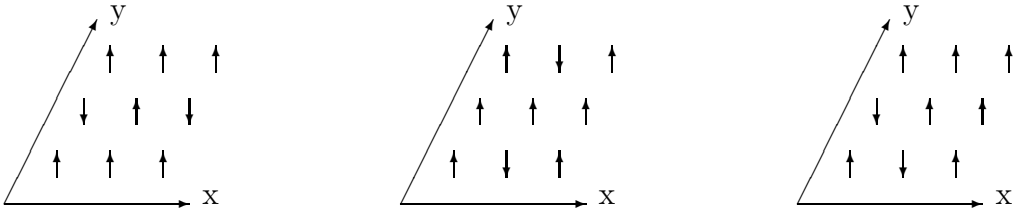
\put(0,20){Fig. 2 Such configurations of spins are possible for the both subsectors.}
\end{picture}

As we see from Fig. 1 the two reversed spins may be neighboring along the Cartesian axes (and hence interacting) only for symmetric states
because according to Eqs. (22) and (24) $a^{(asym)}_{m_1,m_2,n_1,n_2}(k_1,k_2,p_1,p_2)=0$ at $m_2=m_1$ or $n_2=n_1$.

The physical interpretation of the scattering symmetric subsector is the following. According to Eqs. (54) and (62) in the general (not XX) case a scattering two-magnon wave function is a sum of at the minimum $M=3$ terms related to different pairs of two-dimensional wave vectors $(k_1^{(j)},p_1^{(j)})$ and $(k_2^{(j)},p_2^{(j)})$, $j=1,2,3$ (in the XX case $M=2$). According to Eq. (55) an each pair corresponds to the same total energy and total quasimomentum (total wave number). Physically this means that the scattering of two incoming "bare" magnons (for example with wave vectors $(k_1^{(1)},p_1^{(1)})$ and $(k_2^{(1)},p_2^{(1)})$) has a channel resulting in their transformation into at minimum two another pairs (namely the pair $(k_1^{(2)},p_1^{(2)})$, $(k_2^{(2)},p_2^{(2)})$ and the pair $(k_1^{(3)},p_1^{(3)})$, $(k_2^{(3)},p_2^{(3)})$) whose values are restricted {\it only by the energy and quasimomentum conservation laws}. The total energy and quasimomentum of the whole state may be obtained from Eq. (55) by putting here an arbitrary $j$. In other words magnon-magnon interaction results only in the mixing of states with the same energy and quasimomentum.
As it was already mentioned \cite{8} such type of scattering behavior indicates at the same time both ergodicity and non-integrability of the model revealed just in the symmetric two-magnon subsector. Of course the obtained systems of degenerative discrete-diffractive states (54), (62) and (61) are overcomplete and obviously non-orthogonal (moreover they do not contain neither bound no resonant states \cite{10,11,12} which obviously should lie in the symmetric subsector). Nevertheless as it was emphasized by F. Dyson \cite{9} even an overcomplete set of eigenstates may be used for calculation of physical quantities if one obtain for it the corresponding resolution of unity.

\section{Bethe two-magnon states on 3D cubic latice}
On the 3D simple cubic lattice the Heisenberg-Izing Hamiltonian has the form
\begin{eqnarray}
&&\hat H=-\sum_{m,n,r}\Big[\frac{J_1}{2}\Big({\bf S}^-_{m+1,n,r}+{\bf S}^-_{m-1,n,r}\Big){\bf S}^+_{m,n,r}+
\frac{J_2}{2}\Big({\bf S}^-_{m,n+1,r}
+{\bf S}^-_{m,n-1,r}\Big){\bf S}^+_{m,n,r}\nonumber\\
&&+\frac{J_3}{2}\Big({\bf S}^-_{m,n,r+1}+{\bf S}^-_{m,n,r-1}\Big){\bf S}^+_{m,n,r}
+J_{z,1}\Big({\bf S}^z_{m,n,r}{\bf S}^z_{m+1,n,r}-\frac{1}{4}\Big)\nonumber\\
&&+J_{z,2}\Big({\bf S}^z_{m,n,r}{\bf S}^z_{m,n+1,r}-\frac{1}{4}\Big)
+J_{z,3}\Big({\bf S}^z_{m,n,r}{\bf S}^z_{m,n,r+1}-\frac{1}{4}\Big)+\gamma h\Big({\bf S}_{m,n}^z-\frac{1}{2}\Big)\Big],
\end{eqnarray}
where each ${\bf S}_{m,n,r}$ is a spin operator on the site with coordinates $m$, $n$ and $r$.

As in the 2D case we suggest the ferromagnetically polarized ground state
\begin{equation}
|\emptyset\rangle=\prod_{m,n,r=-\infty}^{\infty}\otimes|\uparrow\rangle_{m,n,r},
\end{equation}
similar to (2). As in (3) a one-magnon state is a Bloch wave \cite{9}
\begin{equation}
|k,p,q\rangle=\sum_{m,n,r}{\rm e}^{i(km+pn+qr)}{\bf S}^-_{m,n,r}|\emptyset\rangle,
\end{equation}
with an energy
\begin{equation}
E(k,p,q,h)=E_1(k)+E_2(p)+E_3(q)+\gamma h.
\end{equation}
where as in (5) $E_3(q)=J_{z,3}-J_3\cos{q}$.

A two-magnon state has the form
\begin{equation}
|2\rangle=\sum_{m_1,m_2,n_1,n_2,r_1,r_2}a_{m_1,m_2,n_1,n_2,r_1,r_2}{\bf S}^-_{m_1,n_1,r_1}{\bf S}^-_{m_2,n_2,r_2}|\emptyset\rangle,
\end{equation}
where as in (7) we postulate
\begin{equation}
a_{m_2,m_1,n_2,n_1,r_2,r_1}=a_{m_1,m_2,n_1,n_2,r_1,r_2}.
\end{equation}
Correspondingly
\begin{equation}
\langle2|\tilde2\rangle=\sum_{m_1,m_2,n_1,n_2,r_1,r_2=-\infty}^{\infty}
(1-\delta_{m_1m_2}\delta_{n_1n_2}\delta_{r_1r_2})\bar a_{m_1,m_2,n_1,n_2,r_1,r_2}
\tilde a_{m_1,m_2,n_1,n_2,r_1,r_2}.
\end{equation}

The related ${\rm Schr\ddot odinger}$ equation splits now on four systems
\begin{eqnarray}
&&2(J_{z,1}+J_{z,2}+J_{z,3}+\gamma h)a_{m_1,m_2,n_1,n_2,r_1,r_2}-\sum_{j=\pm1}\Big[
\frac{J_1}{2}\Big(a_{m_1+j,m_2,n_1,n_2,r_2,r_2}\nonumber\\
&&+a_{m_1,m_2+j,n_1,n_2,r_1,r_2}\Big)
+\frac{J_2}{2}\Big(a_{m_1,m_2,n_1+j,n_2,r_1,r_2}+a_{m_1,m_2,n_1,n_2+j,r_1,r_2}\Big)
\nonumber\\
&&+\frac{J_3}{2}\Big(a_{m_1,m_2,n_1,n_2,r_1+j,r_2}+a_{m_1,m_2,n_1,n_2,r_1,r_2+j}\Big)\Big]=
Ea_{m_1,m_2,n_1,n_2,r_1,r_2},
\nonumber\\
&&(m_2-m_1)^2+(n_2-n_1)^2+(r_2-r_1)^2>1,\\
&&(J_{z,1}+2J_{z,2}+2J_{z,3}+2\gamma h)a_{m,m+1,n,n,r,r}-\frac{J_1}{2}\Big(a_{m-1,m+1,n,n,r,r}+a_{m,m+2,n,n,r,r}\Big)\nonumber\\
&&-\sum_{j=\pm1}\Big[\frac{J_2}{2}\Big(a_{m,m+1,n+j,n,r,r}+a_{m,m+1,n,n+j,r,r}\Big)\nonumber\\
&&+\frac{J_3}{2}\Big(a_{m,m+1,n,n,r+j,r}+a_{m,m+1,n,n,r,r+j}\Big)\Big]=Ea_{m,m+1,n,n,r,r},
\\
&&(2J_{z,1}+J_{z,2}+2J_{z,3}+2\gamma h)a_{m,m,n,n+1,r,r}-\frac{J_2}{2}\Big(a_{m,m,n-1,n+1,r,r}+a_{m,m,n,n+2,r,r}\Big)\nonumber\\
&&-\sum_{j=\pm1}\Big[\frac{J_1}{2}\Big(a_{m+j,m,n,n+1,r,r}+a_{m,m+j,n,n+1,r,r}\Big)\nonumber\\
&&+\frac{J_3}{2}\Big(a_{m,m,n,n+1,r+j,r}+a_{m,m,n,n+1,r,r+j}\Big)\Big]=Ea_{m,m,n,n+1,r,r},
\\
&&(2J_{z,1}+2J_{z,2}+J_{z,3}+2\gamma h)a_{m,m,n,n,r,r+1}-\sum_{j=\pm1}\Big[
\frac{J_1}{2}\Big(a_{m+j,m,n,n,r,r+1}+a_{m,m+j,n,n,r,r+1}\Big)\nonumber\\
&&+\frac{J_2}{2}\Big(a_{m,m,n+j,n,r,r+1}+a_{m,m,n,n+j,r,r+1}\Big)\Big]\nonumber\\
&&-\frac{J_3}{2}\Big(a_{m,m,n,n,r-1,r+1}+a_{m,m,n,n,r,r+2}\Big)=Ea_{m,m,n,n,r,r+1}.
\end{eqnarray}

As usual the standard expansion of the wave function into an unphysical region $m_2=m_1$, $n_2=n_1$, $r_2=r_1$ reduces Eqs. (76)-(78) to the form (75) producing as in (16) a system of Bethe conditions
\begin{eqnarray}
&&J_1\Big(a_{m,m,n,n,r,r}+a_{m+1,m+1,n,n,r,r}\Big)=2J_{z,1}a_{m,m+1,n,n,r,r},\nonumber\\
&&J_2\Big(a_{m,m,n,n,r,r}+a_{m,m,n+1,n+1,r,r}\Big)=2J_{z,2}a_{m,m,n,n+1,r,r},\nonumber\\
&&J_3\Big(a_{m,m,n,n,r,r}+a_{m,m,n,n,r+1,r+1}\Big)=2J_{z,3}a_{m,m,n,n,r,r+1}.
\end{eqnarray}

An exponential two-magnon wave function compatible with the symmetry (73) should have the form
\begin{eqnarray}
&&a_{m_1,m_2,n_1,n_2,r_1,r_2}(k_1,k_2,p_1,p_2,q_1,q_2)\nonumber\\
&&=A_0\Big({\rm e}^{i(k_1m_1+k_2m_2+p_1n_1+p_2n_2+q_1r_1+q_2r_2)}
+{\rm e}^{i(k_2m_1+k_1m_2+p_2n_1+p_1n_2+q_2r_1+q_1r_2)}\Big)\nonumber\\
&&+A_1\Big({\rm e}^{i(k_2m_1+k_1m_2+p_1n_1+p_2n_2+q_1r_1+q_2r_2)}
+{\rm e}^{i(k_1m_1+k_2m_2+p_2n_1+p_1n_2+q_2r_1+q_1r_2)}\Big)\nonumber\\
&&+A_2\Big({\rm e}^{i(k_1m_1+k_2m_2+p_2n_1+p_1n_2+q_1r_1+q_2r_2)}
+{\rm e}^{i(k_2m_1+k_1m_2+p_1n_1+p_2n_2+q_2r_1+q_1r_2)}\Big)\nonumber\\
&&+A_3\Big({\rm e}^{i(k_1m_1+k_2m_2+p_1n_1+p_2n_2+q_2r_1+q_1r_2)}
+{\rm e}^{i(k_2m_1+k_1m_2+p_2n_1+p_1n_2+q_1r_1+q_2r_2)}\Big).
\end{eqnarray}
As in the 2D case the wave function (80) solves Eq. (75) giving an energy
\begin{equation}
E(k_1,k_2,p_1,p_2,q_1,q_2,h)=\sum_{j=1}^2\Big(E_1(k_j)+E_2(p_j)+E_3(q_j)\Big)+2\gamma h.
\end{equation}
At the same time a substitution of (80) into (79) results in a system
\begin{eqnarray}
&&(A_0+A_1+A_2+A_3)X^{(1)}(k_1,k_2)=(A_0+A_1+A_2+A_3)X^{(2)}(p_1,p_2)\nonumber\\
&&=(A_0+A_1+A_2+A_3)X^{(3)}(q_1,q_2)=0,
\end{eqnarray}
which is analogous to (19) and solvable only under the condition
\begin{equation}
A_0+A_1+A_2+A_3=0.
\end{equation}

Taking the following basis in the set of solutions of Eq. (83)
\begin{eqnarray}
A_0=A_1=\frac{1}{2\sqrt{2}},\qquad A_2=A_3=-\frac{1}{2\sqrt{2}},\qquad j=1,\nonumber\\
A_0=A_2=\frac{1}{2\sqrt{2}},\qquad A_1=A_3=-\frac{1}{2\sqrt{2}},\qquad j=2,\nonumber\\
A_0=A_3=\frac{1}{2\sqrt{2}},\qquad A_1=A_2=-\frac{1}{2\sqrt{2}},\qquad j=3,
\end{eqnarray}
we readily get the corresponding set of factorized wave functions
\begin{eqnarray}
a^{(asym,1)}_{m_1,m_2,n_1,n_2,r_1,r_2}(k_1,k_2,p_1,p_2,q_1,q_2)
=a^{(s)}_{m_1,m_2}(k_1,k_2)a^{(a)}_{n_1,n_2}(p_1,p_2)a^{(a)}_{r_1,r_2}(q_1,q_2),\nonumber\\
a^{(asym,2)}_{m_1,m_2,n_1,n_2,r_1,r_2}(k_1,k_2,p_1,p_2,q_1,q_2)
=a^{(a)}_{m_1,m_2}(k_1,k_2)a^{(s)}_{n_1,n_2}(p_1,p_2)a^{(a)}_{r_1,r_2}(q_1,q_2),\nonumber\\
a^{(asym,3)}_{m_1,m_2,n_1,n_2,r_1,r_2}(k_1,k_2,p_1,p_2,q_1,q_2)
=a^{(a)}_{m_1,m_2}(k_1,k_2)a^{(a)}_{n_1,n_2}(p_1,p_2)a^{(s)}_{r_1,r_2}(q_1,q_2),
\end{eqnarray}
which are the 3D analogs of the 2D antisymmetric ones (22). According to Eqs. (28), (52) and their counterparts
\begin{equation}
P^{(asym)}a^{(s)}(k_1,k_2)=P^{(sym)}a^{(a)}(k_1,k_2)=0,
\end{equation}
the wave functions (85) satisfy the following analogs of Eq. (29)
\begin{equation}
{\cal P}^{(asym,j)}a^{(asym,l)}=\delta_{jl}a^{(asym,j)},\qquad j,l=1,2,3,
\end{equation}
where
\begin{eqnarray}
{\cal P}^{(asym,1)}=P^{(sym)}\otimes P^{(asym)}\otimes P^{(asym)},\nonumber\\
{\cal P}^{(asym,2)}=P^{(asym)}\otimes P^{(sym)}\otimes P^{(asym)},\nonumber\\
{\cal P}^{(asym,3)}=P^{(asym)}\otimes P^{(asym)}\otimes P^{(sym)}.
\end{eqnarray}

According to Eqs. (32) and (63) we may prove formulas analogous to (33)
\begin{eqnarray}
&&a^{(asym,1)}(k_1,k_2,p_1,p_2,q_1,q_2)=a^{(asym,1)}(k_2,k_1,p_1,p_2,q_1,q_2)\nonumber\\
&&=-a^{(asym,1)}(k_1,k_2,p_2,p_1,q_1,q_2)
=-a^{(asym,1)}(k_1,k_2,p_1,p_2,q_2,q_1),\nonumber\\
&&a^{(asym,2)}(k_1,k_2,p_1,p_2,q_1,q_2)=-a^{(asym,2)}(k_2,k_1,p_1,p_2,q_1,q_2)\nonumber\\
&&=a^{(asym,2)}(k_1,k_2,p_2,p_1,q_1,q_2)
=-a^{(asym,2)}(k_1,k_2,p_1,p_2,q_2,q_1),\nonumber\\
&&a^{(asym,3)}(k_1,k_2,p_1,p_2,q_1,q_2)=-a^{(asym,3)}(k_2,k_1,p_1,p_2,q_1,q_2)\nonumber\\
&&=-a^{(asym,3)}(k_1,k_2,p_2,p_1,q_1,q_2)
=a^{(asym,3)}(k_1,k_2,p_1,p_2,q_2,q_1),
\end{eqnarray}
from which follows that as in (34) we may put
\begin{equation}
0\leq k_1<k_2<2\pi,\qquad0\leq p_1<p_2<2\pi\qquad0\leq q_1<q_2<2\pi.
\end{equation}

In the same manner as it was done for Eqs. (35) and (37) one may readily prove that
\begin{eqnarray}
&&\sum_{m_1,m_2=-\infty}^{\infty}\bar a^{(s)}_{m_1,m_2}(k_1,k_2)a^{(s)}_{m_1,m_2}(\tilde k_1,\tilde k_2)=(2\pi)^2
\delta(\tilde k_1-k_1)\delta(\tilde k_2-k_2),\\
&&\frac{1}{(2\pi)^2}\int_0^{2\pi}dk_2\int_0^{k_2}dk_1
\bar a^{(s)}_{m_1,m_2}(k_1,k_2)a^{(s)}_{\tilde m_1,\tilde m_2}(k_1,k_2)=P^{(sym)}_{m_1,m_2,\tilde m_1,\tilde m_2}.
\end{eqnarray}
Now Eqs. (35), (37), (91), (92) and (85) result in the following analog of Eqs. (36) and (39)
\begin{eqnarray}
&&\langle\tilde k_1,\tilde k_2,\tilde p_1,\tilde p_2,\tilde q_1,\tilde q_2,asym,\tilde j|k_1,k_2,p_1,p_2,q_1,q_2,asym,j\rangle\nonumber\\
&&=(2\pi)^6\delta_{j\tilde j}\delta(\tilde k_1-k_1)\delta(\tilde k_2-k_2)\delta(\tilde p_1-p_1)\delta(\tilde p_2-p_2)\delta(\tilde q_1-q_1)\delta(\tilde q_2-q_2),\\
&&\frac{1}{(2\pi)^6}\int_0^{2\pi}dk_2\int_0^{k_2}dk_1\int_0^{2\pi}dp_2\int_0^{p_2}dp_1
\int_0^{2\pi}dq_2\int_0^{q_2}dq_1\nonumber\\
&&\cdot|k_1,k_2,p_1,p_2,q_1,q_2,asym,j\rangle\langle k_1,k_2,p_1,p_2,q_1,q_2,asym,j|={\cal P}^{(asym,j)}.
\end{eqnarray}

For a wave function related to a finite periodic $N_x\times N_y\times N_z$ lattice one may readily obtain equations analogous to (40)-(43). Namely in the 3D case
\begin{equation}
N^{(asym)}=\frac{3N_xN_yN_z(N_x-1)(N_y-1)(N_z-1)}{8},\qquad N^{(tot)}=\frac{N_xN_yN_z(N_xN_yN_z-1)}{2}.
\end{equation}
Hence for $N_x,N_y,N_z\gg1$ one has from (95)
\begin{equation}
N^{(asym)}\approx\frac{3N^{(tot)}}{4}.
\end{equation}

The 3D analog of the 2D symmetric subsector corresponds to the wave functions
\begin{eqnarray}
&&a^{(DDD,sym)}_{m_1,m_2,n_1,n_2,r_1,r_2}(\{k_1^{(1)},k_2^{(1)},p_1^{(1)},p_2^{(1)},q_1^{(1)},q_2^{(1)}\},\dots,
\{k_1^{(M)},k_2^{(M)},p_1^{(M)},p_2^{(M)},q_1^{(M)},q_2^{(M)}\})\nonumber\\
&&=\sum_{j=1}^MB_ja^{(sym)}_{m_1,m_2,n_1,n_2,r_1,r_2}(k_1^{(j)},k_2^{(j)},p_1^{(j)},p_2^{(j)},q_1^{(j)},q_2^{(j)}),
\end{eqnarray}
where according to the conservation laws
\begin{eqnarray}
&&{\rm e}^{i(k_1^{(j)}+k_2^{(j)})}={\rm e}^{ik},\qquad{\rm e}^{i(p_1^{(j)}+p_2^{(j)})}={\rm e}^{ip},\qquad{\rm e}^{i(q_1^{(j)}+q_2^{(j)})}={\rm e}^{iq}\nonumber\\
&&E_1(k_1^{(j)})+E_1(k_2^{(j)})+E_2(p_1^{(j)})+E_2(p_2^{(j)})+E_3(q_1^{(j)})+
E_3(q_2^{(j)})=E-2\gamma h.
\end{eqnarray}
Here $k$, $p$ and $q$ are fixed numbers related to total wave numbers along the Cartesian axes.
A substitution of (97) and (98) reduces the system (79) to the form
\begin{equation}
\sum_{j=1}^MB_jX^{(1)}(k_1^{(j)},k_2^{(j)})=0,\qquad
\sum_{j=1}^MB_jX^{(2)}(p_1^{(j)},p_2^{(j)})=0,
\qquad\sum_{j=1}^MB_jX^{(3)}(q_1^{(j)},q_2^{(j)})=0.
\end{equation}

In general case the DDD solution of the system (99) exist already at $M=4$ and has the form
\begin{equation}
B_j=\sum_{l,m,n=1}^4\varepsilon_{jlmn}X^{(1)}(k_1^{(l)},k_2^{(l)})
X^{(2)}(p_1^{(m)},p_2^{(m)})X^{(3)}(q_1^{(n)},q_2^{(n)}).
\end{equation}
For the $XX$ model when additionally to Eq. (58) one has
\begin{equation}
J_{3,z}=0\Longrightarrow X^{(3)}(q_1^{(j)},q_2^{(j)})=J_3\Big(1+{\rm e}^{iq}\Big),
\end{equation}
the system (99) again reduces to Eq. (59) producing the $M=2$ solution
\begin{eqnarray}
&&a^{(DDD,sym,XX)}_{m_1,m_2,n_1,n_2,r_1,r_2}
(k_1,k_2,p_1,p_2,q_1,q_2,\tilde k_1,\tilde k_2,\tilde p_1,\tilde p_2,\tilde q_1,\tilde q_2)
\nonumber\\
&&=a^{(s)}_{m_1,m_2}(k_1,k_2)a^{(s)}_{n_1,n_2}(p_1,p_2)a^{(s)}_{r_1,r_2}(q_1,q_2)
-a^{(s)}_{m_1,m_2}(\tilde k_1,\tilde k_2)a^{(s)}_{n_1,n_2}(\tilde p_1,\tilde p_2)
a^{(s)}_{r_1,r_2}(\tilde q_1,\tilde q_2).\qquad\,
\end{eqnarray}
analogous to (61).

\section{Some remarks on DDD BA approach}

\subsection{Splitting of two-magnon sector on subsectors}

In the present paper we have utilized the splitting of the two-magnon sectors on the antisymmetric and symmetric subsectors. Do the antisymmetric and symmetric wavefunctions exhaust all the states in the two-magnon sector? The answer is "yes". Really a 2D two-magnon wave function $a_{m_1,m_2,n_1,n_2}$ which satisfy the symmetry condition (7) may be represented as a sum of its antisymmetric and symmetric components
\begin{equation}
a_{m_1,m_2,n_1,n_2}=a^{(asym)}_{m_1,m_2,n_1,n_2}+a^{(sym)}_{m_1,m_2,n_1,n_2},
\end{equation}
where
\begin{eqnarray}
&&a^{(asym)}_{m_1,m_2,n_1,n_2}=\frac{1}{2}\Big(a_{m_1,m_2,n_1,n_2}-a_{m_1,m_2,n_2,n_1}\Big),\nonumber\\
&&a^{(sym)}_{m_1,m_2,n_1,n_2}=\frac{1}{2}\Big(a_{m_1,m_2,n_1,n_2}+a_{m_1,m_2,n_2,n_1}\Big).
\end{eqnarray}
Then a substitution of (103) into the ${\rm Schr\ddot odinger}$ equation (13)-(15) results in two separate systems for both the counterparts.

In 3D the corresponding result follows from an expansion
\begin{equation}
a_{m_1,m_2,n_1,n_2,r_1,r_2}=\sum_{j=1}^3a^{(asym,j)}_{m_1,m_2,n_1,n_2,r_1,r_2}+a^{(sym)}_{m_1,m_2,n_1,n_2,r_1,r_2},
\end{equation}
where
\begin{eqnarray}
&&a^{(asym,1)}_{m_1,m_2,n_1,n_2,r_1,r_2}=\frac{1}{4}\Big(a_{m_1,m_2,n_1,n_2,r_1,r_2}-a_{m_1,m_2,n_2,n_1,r_1,r_2}\nonumber\\
&&-a_{m_1,m_2,n_1,n_2,r_2,r_1}+a_{m_1,m_2,n_2,n_1,r_2,r_1}\Big),\nonumber\\
&&a^{(asym,2)}_{m_1,m_2,n_1,n_2,r_1,r_2}=\frac{1}{4}\Big(a_{m_1,m_2,n_1,n_2,r_1,r_2}-a_{m_2,m_1,n_1,n_2,r_1,r_2}\nonumber\\
&&-a_{m_1,m_2,n_1,n_2,r_2,r_1}+a_{m_2,m_1,n_1,n_2,r_2,r_1}\Big),\nonumber\\
&&a^{(asym,3)}_{m_1,m_2,n_1,n_2,r_1,r_2}=\frac{1}{4}\Big(a_{m_1,m_2,n_1,n_2,r_1,r_2}-a_{m_2,m_1,n_1,n_2,r_1,r_2}\nonumber\\
&&-a_{m_1,m_2,n_2,n_1,r_1,r_2}+a_{m_2,m_1,n_2,n_1,r_1,r_2}\Big),
\end{eqnarray}
or according to (73)
\begin{eqnarray}
&&a^{(asym,1)}_{m_1,m_2,n_1,n_2,r_1,r_2}=\frac{1}{4}\Big(a_{m_1,m_2,n_1,n_2,r_1,r_2}-a_{m_1,m_2,n_2,n_1,r_1,r_2}\nonumber\\
&&-a_{m_1,m_2,n_1,n_2,r_2,r_1}+a_{m_2,m_1,n_1,n_2,r_1,r_2}\Big),\nonumber\\
&&a^{(asym,2)}_{m_1,m_2,n_1,n_2,r_1,r_2}=\frac{1}{4}\Big(a_{m_1,m_2,n_1,n_2,r_1,r_2}-a_{m_2,m_1,n_1,n_2,r_1,r_2}\nonumber\\
&&-a_{m_1,m_2,n_1,n_2,r_2,r_1}+a_{m_1,m_2,n_2,n_1,r_1,r_2}\Big),\nonumber\\
&&a^{(asym,3)}_{m_1,m_2,n_1,n_2,r_1,r_2}=\frac{1}{4}\Big(a_{m_1,m_2,n_1,n_2,r_1,r_2}-a_{m_2,m_1,n_1,n_2,r_1,r_2}\nonumber\\
&&-a_{m_1,m_2,n_2,n_1,r_1,r_2}+a_{m_1,m_2,n_1,n_2,r_2,r_1}\Big),
\end{eqnarray}
and
\begin{eqnarray}
&&a^{(sym)}_{m_1,m_2,n_1,n_2,r_1,r_2}=\frac{1}{4}\Big(a_{m_1,m_2,n_1,n_2,r_1,r_2}+a_{m_2,m_1,n_1,n_2,r_1,r_2}\nonumber\\
&&+a_{m_1,m_2,n_2,n_1,r_1,r_2}+a_{m_1,m_2,n_1,n_2,r_2,r_1}\Big).
\end{eqnarray}

\subsection{Special role of the 1D case}

As it was shown above for $D\geq2$ in the general (not $XX$!) case the symmetric wave function has DDD Bethe form with $M=D+1$. This readily follows from the fact that the magnon-magnon scattering produces $D$ Bethe conditions (see Eqs. (16), (79) and (148) below) on amplitudes which in their turn are defined up to a common factor. However an extrapolation of this result on the $D=1$ case is incorrect. Really the $D=1$ system
\begin{equation}
\hat H=-\sum_{n=-\infty}^{\infty}\Big[\frac{J}{2}\Big({\bf S}^-_{n-1}+{\bf S}^-_{n+1}\Big){\bf S}^+_n
+J_z\Big({\bf S}^z_n{\bf S}^z_{n+1},-\frac{1}{4}\Big)+\gamma h\Big({\bf S}_n^z-\frac{1}{2}\Big)\Big],
\end{equation}
is integrable and hence corresponds to $M=1$. In order to clarify this special feature of the 1D case let us consider
the representation of a two-magnon state
\begin{equation}
|2\rangle=\sum_{n_1<n_2}a_{n_1,n_2}{\bf S}^-_{n_1}{\bf S}^-_{n_2}|\emptyset\rangle,
\end{equation}
which utilizes the ordering of the magnon coordinates ($n_1<n_2$) and hence makes unnecessary the symmetry condition $a_{n_2,n_1}=a_{n_1,n_2}$ (the 1D analog of Eqs. (7) or (73)).
The corresponding ${\rm Schr\ddot odinger}$ equation
\begin{eqnarray}
&&2(J_z+\gamma h)a_{n_1,n_2}-\frac{J}{2}\sum_{l=\pm1}
\Big(a_{n_1+l,n_2}+a_{n_1,n_2+l}\Big)
=Ea_{n_1,n_2},\quad n_2-n_1>1,\\
&&(J_z+2\gamma h)a_{n,n+1}
-\frac{J}{2}\Big(a_{n-1,n+1}+a_{n,n+2}\Big)=Ea_{n,n+1}.
\end{eqnarray}
has for $n_2-n_1>1$ (Eq. (111)) a two-parametric solution
\begin{equation}
a_{n_1,n_2}=A{\rm e}^{i(k_1n_1+k_2n_2)}-\tilde A{\rm e}^{i(k_2n_1+k_1n_2)},
\end{equation}
where both the amplitudes $A$ and $\tilde A$ are {\it independent}.

An equivalent to Eq. (112) Bethe condition
\begin{equation}
J\Big(a_{n,n}+a_{n+1,n+1}\Big)=2J_za_{n,n+1},
\end{equation}
(analogous to Eqs. (16) and (79)) gives up to a constant factor
\begin{equation}
A=A(k_1,k_2),\qquad\tilde A=A(k_2,k_1),
\end{equation}
where
\begin{equation}
A(k_1,k_2)=J\cos{\frac{k_1+k_2}{2}}-J_z{\rm e}^{i(k_1-k_2)/2}.
\end{equation}
According to (115), (116) the solution (113) is neither antisymmetric (it however antisymmetric in the $XX$ case when $J_z=0$) no symmetric.

Of course one may represent the two-magnon state (110) in an equivalent "symmetric" form
\begin{equation}
|2\rangle=\sum_{n_1,n_2}a_{n_1,n_2}^{(sym)}{\bf S}^-_{n_1}{\bf S}^-_{n_2}|\emptyset\rangle,\qquad a_{n_2,n_1}^{(sym)}=a_{n_1,n_2}^{(sym)},
\end{equation}
similar to (6) and (72), where
\begin{equation}
a_{n_1,n_2}^{(sym)}={\rm e}^{i(k_1+k_2)(n_1+n_2)/2}\Big(A{\rm e}^{i(k_2-k_1)|n_1-n_2|}-
\tilde A{\rm e}^{i(k_1-k_2)|n_1-n_2|}\Big).
\end{equation}
Since both the exponents in this representations are invariant under the symmetry $n_1\leftrightarrow n_2$, both the amplitudes again are independent before a substitution of (118) into (114).

Following Eq. (118) one may suggest a 2D two-magnon {\it symmetric} wave function in the form
\begin{eqnarray}
&&a^{(sym)}_{m_1,m_2,n_1,n_2}(k,\kappa,p,\rho)={\rm e}^{i[k(m_1+m_2)/2+p(n_1+n_2)/2]}\Big(A_1{\rm e}^{i(\kappa|m_1-m_2|+\rho|n_1-n_2|)}\nonumber\\
&&+A_2{\rm e}^{i(\kappa|m_1-m_2|-\rho|n_1-n_2|)}+A_3{\rm e}^{i(-\kappa|m_1-m_2|+\rho|n_1-n_2|)}+A_4{\rm e}^{-i(\kappa|m_1-m_2|+\rho|n_1-n_2|)}\Big),
\qquad
\end{eqnarray}
which obviously satisfy the symmetry (45). Since ${\rm e}^{i\pi|m_1-m_2|}={\rm e}^{i\pi(m_1+m_2)}$ we may put in (119)
\begin{equation}
0\leq\kappa<\pi,\qquad0\leq\rho<\pi.
\end{equation}

The wave function (119) satisfies Eq. (13) for $(m_1-m_2)(n_1-n_2)\neq0$ giving an energy
\begin{equation}
E(k,\kappa,p,\rho,h)=E_1\Big(\frac{k}{2}-\kappa\Big)+E_1\Big(\frac{k}{2}+\kappa\Big)+E_2\Big(\frac{p}{2}-\rho\Big)+
E_2\Big(\frac{p}{2}+\rho\Big)+2\gamma h.
\end{equation}
Being depending on the four amplitudes ($A_j$, $j=1,\dots,4$), the wave function (119) seems to be more general than (50).
However an account of (13) at $n_1=n_2$, $|m_1-m_2|>1$ and $m_1=m_2$, $|n_1-n_2|>1$ results in the following system of conditions
\begin{eqnarray}
&&(A_1-A_2)\sin{\rho}=(A_3-A_4)\sin{\rho}=0,\\
&&(A_1-A_3)\sin{\kappa}=(A_2-A_4)\sin{\kappa}=0.
\end{eqnarray}
Really, taking $n_1=n_2$ and $|m_1-m_2|>1$ and using (45) and (121) one readily reduces Eq. (13) to
\begin{eqnarray}
&&\sum_{l=\pm1}\Big[\frac{J_1}{2}\Big(a^{(sym)}_{m_1+l,m_2,n,n}+a^{(sym)}_{m_1,m_2+l,n,n}\Big)+J_2a^{(sym)}_{m_1,m_2,n,n+l}\Big]\nonumber\\
&&=2\Big(J_1\cos{\frac{k}{2}}\cos{\kappa}+J_2\cos{\frac{p}{2}}\cos{\rho}\Big)a^{(sym)}_{m_1,m_2,n,n}.
\end{eqnarray}
Now it may be readily proved that if $a^{(sym)}_{m_1,m_2,n,n}$ is given by Eq. (119), then Eq. (124) at $|m_1-m_2|>1$ takes the form

\begin{equation}
\sum_{l=\pm1}a^{(sym)}_{m_1,m_2,n,n+l}(k,\kappa,p,\rho)=2\cos{\frac{p}{2}}\cos{\rho}a^{(sym)}_{m_1,m_2,n,n}(k,\kappa,p,\rho),
\end{equation}
which after some calculations turns into Eq. (122). In the same manner taking $m_1=m_2$ and $|n_1-n_2|>1$ one readily gets Eq. (123)
The system (120), (122), (123) has the following set of solutions
\begin{eqnarray}
&&A_1=A_2=A_3=A_4,\\
&&A_1=A_2,\qquad A_3=A_4,\qquad\kappa=0,\\
&&A_1=A_3,\qquad A_2=A_4,\qquad\rho=0,\\
&&\kappa=\rho=0.
\end{eqnarray}
A simple analysis shows that under each of the conditions (126)-(129) the wave function (119) is equivalent to (50). Comparing now the calculations based on symmetric 1D and 2D wave functions (118) and (119), we see that ratios of the coefficients $A_j$ ($j=1,\dots4$) in (119) are defined {\it before} utilizing the Bethe conditions (16). At the same time the ratio of the coefficients $A$ and $\tilde A$ in (118) may be obtained only from the Bethe condition (114).
The above argumentation shows that the difference between 1D and 2D cases is more deep than between the ones related to $D$ and $D+1$ at $D>1$.

\subsection{DDD BA intuition}

What is an origin of the difficulties inherent in the traditional ($M=1$) BA? Let us first study a two-magnon case. When the magnons are away from each other the ${\rm Schr\ddot odinger}$ equation is automatically satisfied by exponential wave functions related to free motion (for example both the exponents in Eq. (113) separately satisfy Eq. (111)). Utilizing these exponents we construct a wave function which will be called the zero order pig. The latter should satisfy the three following conditions. First of all, it should depend on a minimal set of wave numbers, namely on $D+D=2D$ projections of individual quasimomentums on the Cartesian axes. Secondly, it should be a linear combination with free coefficients (amplitudes) of the maximal set of exponents related to all possible permutations of the wave numbers. Finitely it should of course satisfy all the postulated symmetries (in our case (7) and (73)). In the present paper such zero order pigs are (17), (80) and (113). On the next stage we substitute the zero order pig into the system of Bethe conditions. Though originally a Bethe condition is an {\it infinite} system of equations on the components of wave function (Eq. (114) depends on the parameter $-\infty<n<\infty$) the above substitution yields a {\it single} linear equation on the amplitudes. This remarkable reduction follows from the translation invariance of the lattice according to which all dependence on the scattering point is concentrated in the common exponential factor (in 2D it is ${\rm e}^{i[(k_1+k_2)m+(p_1+p_2)n]}$, see Eq. (19)) which may be canceled.

If the obtained finite linear system of equations (number of equations $=$ number of Bethe conditions) on the amplitudes is solvable (as in $D=1$) then the traditional BA is successful, if not (as for the 2D symmetric subsector) we have to use the DDD BA taking an extended zero order pig. The latter is a sum of the $M$ initial (non-extended) ones and depends on $M$ different sets of wave numbers which however should have {\it equal total energies and equal total quasimomentums}. The extended pig obviously satisfies the ${\rm Schr\ddot odinger}$ equation related to free motion of separated magnons, as well as all the necessary symmetries. Moreover (and this is the keystone of the whole approach) its substitution into any of the Bethe conditions again yields a single equation (obtained after a cancelation of the {\it common} factor which express the spatial dependence). As a result we get the same number of linear equations ($=$ the number of Bethe conditions) on an extended set of amplitudes. For rather big $M$ this linear system of equations will be solvable. Since this argumentation may be applied to each $D$-dimensional {\it translationally invariant} lattice, we have not found any reasons for two-magnon DDD BA to fail in all these cases.

Let us turn to the 1D three-magnon problem for which a zero order pig of wave function has the form
\begin{equation}
a_{n_1,n_2,n_3}=\sum_{j,l,m=1}^3\varepsilon_{jlm}A_{jlm}{\rm e}^{i(k_jn_1+k_ln_2+k_mn_3)},\qquad n_1<n_2<n_3,
\end{equation}
where $\varepsilon_{jlm}$ is the Levy-Civita tensor.
For some models the wave function contain additional terms \cite{8}, however the case (130) is rather representative \cite{16}, \cite{17}. In general case the six amplitudes $A_{jlm}$ are vectors in the space of internal degrees of freedom (usually electron \cite{16} or atomic \cite{17} spin polarization).

At $n_2-n_1>1$, $n_3-n_2>1$ all the exponents in (130) separately satisfy the corresponding free ${\rm Schr\ddot odinger}$ equation. A collision between the particles 1 and 2 occurs at $n_2=n_1+1$ and results in the following {\it infinite} system of linear conditions \cite{17}
\begin{equation}
\sum_{j,l,m=1}^3\varepsilon_{jlm}\Big(G(k_l,k_j)A_{jlm}\Big){\rm e}^{i((k_j+k_l)n_1+k_mn_3)}=0,
\end{equation}
where an explicit form of the matrix $G(k,\tilde k)$ depends on the model.
We see that the exponential factors in (131) are not all equal to each other because they do not depend on the total quasimomentum. However they may be subdivided on equal pairs related to permutations $k_j\leftrightarrow k_l$. This results on the following ({\it finite!}) system of three equations
\begin{equation}
G(k_l,k_j)A_{jlm}=G(k_j,k_l)A_{ljm},\qquad j,l,m=1,2,3,\quad\varepsilon_{jlm}\neq0.
\end{equation}
An analogous system
\begin{equation}
G(k_m,k_l)A_{jlm}=G(k_l,k_m)A_{jml},\qquad j,l,m=1,2,3,\quad\varepsilon_{jlm}\neq0,
\end{equation}
may be obtained after an account of scattering between the particles 2 and 3.

It may be readily shown \cite{16,17,18} that the joint system (132), (133) is solvable only under the Yang-Baxter equation
\begin{equation}
S_{12}(k_l,k_m)S_{23}(k_j,k_m)S_{12}(k_j,k_l)=S_{23}(k_j,k_l)S_{12}(k_j,k_m)S_{23}(k_l,k_m),
\end{equation}
on the $S$-matrix $S(k,\tilde k)=G^{-1}(k,\tilde k)G(\tilde k,k)$.
Here in (134)
\begin{equation}
S_{12}(k,\tilde k)=S(k,\tilde k)\otimes I,\quad S_{23}(k,\tilde k)=I\otimes S(k,\tilde k),
\end{equation}
where $I$ is the identity matrix of the same dimension as the amplitudes in (130).

Can we apply the DDD BA if Eq. (134) fails and the system (132), (133) is unsolvable? For most of the models the answer is negative. Really if we add to the zero order pig (130) an analogous new term
\begin{equation}
\tilde a_{n_1,n_2,n_3}=\sum_{j,l,m=1}^3\varepsilon_{jlm}\tilde A_{jlm}{\rm e}^{i(\tilde k_jn_1+\tilde k_ln_2+\tilde k_mn_3)},\qquad n_1<n_2<n_3,
\end{equation}
with
\begin{eqnarray}
&&\tilde k_1+\tilde k_2+\tilde k_3=k_1+k_2+k_3,\nonumber\\
&&E(\tilde k_1)+E(\tilde k_2)+E(\tilde k_3)=E(k_1)+E(k_2)+E(k_3),
\end{eqnarray}
(where $E(k)$ is the corresponding magnon energy), then Eq. (131) will turn into
\begin{equation}
\sum_{j,l,m=1}^3\varepsilon_{jlm}\Big[\Big(G(k_l,k_j)A_{jlm}\Big){\rm e}^{i((k_j+k_l)n_1+k_mn_3)}+\Big(G(\tilde k_l,\tilde k_j)\tilde A_{jlm}\Big){\rm e}^{i((\tilde k_j+\tilde k_l)n_1+\tilde k_mn_3)}\Big]=0.
\end{equation}
If all the wave numbers in the set $\{\tilde k_1,\tilde k_2,\tilde k_3\}$ are different from the wave numbers in the set $\{k_1,k_2,k_3\}$ then such an addition results only in a new insolvable system on $\{\tilde k_1,\tilde k_2,\tilde k_3\}$ and $\tilde A_{jlm}$ similar to (132), (133).
If however for example
\begin{equation}
\tilde k_3=k_3
\end{equation}
then Eqs. (131) and (138) result in the connected system
\begin{eqnarray}
&&G(k_2,k_3)A_{321}=G(k_3,k_2)A_{231},\qquad G(k_1,k_3)A_{312}=G(k_3,k_1)A_{132},\nonumber\\
&&G(\tilde k_2,k_3)\tilde A_{321}=G(k_3,\tilde k_2)\tilde A_{131},\qquad G(\tilde k_1,k_3)\tilde A_{312}=G(k_3,\tilde k_1)\tilde A_{132},\nonumber\\
&&G(k_1,k_2)A_{213}+G(\tilde k_1,\tilde k_2)\tilde A_{213}=G(k_2,k_1)A_{123}+G(\tilde k_2,\tilde k_1)\tilde A_{123},
\end{eqnarray}
supplemented by its counterpart
\begin{eqnarray}
&&G(k_2,k_3)A_{132}=G(k_3,k_2)A_{123},\qquad G(k_3,k_1)A_{213}=G(k_1,k_3)A_{231},\nonumber\\
&&G(\tilde k_2,k_3)\tilde A_{132}=G(k_3,\tilde k_2)\tilde A_{123},\qquad G(k_3,\tilde k_1)\tilde A_{213}=G(\tilde k_1,k_3)\tilde A_{231},\nonumber\\
&&G(k_1,k_2)A_{321}+G(\tilde k_1,\tilde k_2)\tilde A_{321}=G(k_2,k_1)A_{312}+G(\tilde k_2,\tilde k_1)\tilde A_{312},
\end{eqnarray}
related to scattering between the particles 2 and 3 (see Eq. (133)). Unlike the joint system (132), (133) for solvability of the system (140), (141) the Eq. (134) is unnecessary. However now there is a new obstacle. Under the condition (139) the system (137) reduces to
\begin{equation}
\tilde k_1+\tilde k_2=k_1+k_2,\qquad
E(\tilde k_1)+E(\tilde k_2)=E(k_1)+E(k_2).
\end{equation}
So if we want the system (140), (141) to be really new with respect to (132), (133) then for given $k_1$ and $k_2$ the system (142) should have an additional solution relatively to the two elementary ones $\tilde k_1=k_1$, $\tilde k_2=k_2$ and $\tilde k_1=k_2$, $\tilde k_2=k_1$. Let us notice that though this scenario seems to be reliable its concrete realization is unknown for the author.

If the Yang-Baxter equation (134) is not satisfied or the system (142) for given $k_1$ and $k_2$ has only elementary solutions then the DDD BA is ineffective and the wave function should have a very complex diffractive form \cite{1} (see also Eq. (8) in \cite{8}). In the opposite case solving Eqs. (132), (133) or (140), (141) we shall get the next, {\it first order}, pig which then should be substituted into the ${\rm Schr\ddot odinger}$ equation related to the three-magnon collision ($n_1=n_2-1$, $n_3=n_2+1$). Since in the latter case the dependence of the corresponding  Bethe conditions on the collision point reduces to the factor ${\rm e}^{(k_1+k_2+k_3)n_2}$ it may be successfully treated within the DDD BA machinery.

Finitely we may conclude that the DDD BA should be effective for evaluation of two-magnon states for Heisenberg-Ising ferromagnet on arbitrary translation invariant D-dimensional infinite lattices. Very likely that it is so also for some other models whose ground states have the tensor-product form similar to (2) and (69). At the same time an application of DDD BA to a three-magnon sector is a more delicate problem. First of all we should obtain a first order pig which accounts all pair collisions in the presence of the third remote magnon. As it was shown in Sect. 6 at $D=1$ an existence of such a pig is guaranteed by Eq. (134) \cite{18}. If it is not satisfied then in principle the DDD BA may be applied only if for each triple of wave numbers $\{k_1,k_2,k_3\}$ corresponds a {\it different} (not obtained by a permutations of wave numbers) triple $\{\tilde k_1,\tilde k_2,\tilde k_3\}$ which satisfy Eqs. (139) and (142) (as it was already mentioned any model for which such alternative is realized is unknown for the author). Having now the first order pig we may study the Bethe conditions related to pure three-magnon collisions. For integrable 1D models \cite{1,2,3} this first order pig completely solves the three-magnon problem. At the same time for the 1D $S=1$ ferromagnet (for which Eq. (134) is satisfied automatically because the matrix $S(k,\tilde k)$ is one-dimensional) a three-magnon scattering produces an additional Bethe condition (on the first order pig) which may be solved only within the DDD BA machinery \cite{8}. Physically this phenomena originates from an existence of two-magnon resonance state related to transitions of two neighboring spins with polarizations ${\bf S}^z=0$ into a single one with polarization ${\bf S}^z=-1$ (polarization ${\bf S}^z=1$ corresponds to the ground state). For the Heisenberg-Ising ferromagnet at {\it any dimensions} an excited site has the single polarization ${\bf S}^z=-1/2$ and three-magnon collisions do not yield additional Bethe conditions. Hence an account of all pair collisions in the presence of the third remote magnon is sufficient for solution of the three-magnon problem.

Is there a $(D>1)$-dimensional generalization of the condition (134) under which the three-magnon problem is solvable within DDD BA? All that we have at the present time is the so called tetrahedron equation \cite{19,20} on a three-magnon $S$-matrix, which guarantees solvability of the {\it four}-magnon problem. So we can not give an answer on the suggested question. Nevertheless, it seems that for our model this is non essential. Really both Eq. (134) and the tetrahedron equation imply that the system has internal degrees of freedom which is not so for the Heisenberg-Ising ferromagnet in {\it any dimensions!} Moreover, as it will be shown in Sect. 8 both the antisymmetric (for the general XXZ model) and symmetric subsectors (for the XX model) may be readily extracted from {\it all} $Q$-magnon sectors in any dimensions. Of course for $Q>2$ they do not exhaust all the states, but this is an another problem.

\section{Two-magnon wave function at $D>3$}

The representations obtained previously for 2D and 3D two-magnon wave functions of the Heisenberg-Ising ferromagnet may be generalized to arbitrary $D$ where the ${\rm Schr\ddot odinger}$ equation related to the state
\begin{equation}
|2\rangle=\sum_{n^{(1)}_1,n^{(1)}_2,\dots,n^{(D)}_1,n^{(D)}_2}a_{n^{(1)}_1,n^{(1)}_2,\dots,n^{(D)}_1,n^{(D)}_2}
{\bf S}^-_{n^{(1)}_1,\dots n^{(D)}_1}{\bf S}^-_{n^{(1)}_2,\dots n^{(D)}_2}|\emptyset\rangle,
\end{equation}
has the form
\begin{eqnarray}
&&\Big(2\sum_{m=1}^DJ_{z,m}+2\gamma h-E\Big)a_{n^{(1)}_1,n^{(1)}_2,\dots,n^{(D)}_1,n^{(D)}_2}
-\sum_{m=1}^D\frac{J_m}{2}\sum_{j=\pm1}\Big(a_{\dots,n^{(m)}_1+j,n^{(m)}_2,\dots}\nonumber\\
&&+a_{\dots,n^{(m)}_1,n^{(m)}_2+j,\dots}\Big)=0,
\qquad\sum_{m=1}^D\Big(n^{(m)}_2-n^{(m)}_1\Big)^2>1,\\
&&\Big(2\sum_{m=1}^DJ_{z,m}-J_{z,l}+2\gamma h-E\Big)a_{n^{(1)},n^{(1)},\dots,n^{(l)},n^{(l)}+1,\dots,n^{(D)},n^{(D)}}\nonumber\\
&&-\frac{J_l}{2}\Big(a_{n^{(1)},n^{(1)},\dots,n^{(l)}-1,n^{(l)}+1,\dots,n^{(D)},n^{(D)}}
+a_{n^{(1)},n^{(1)},\dots,n^{(l)},n^{(l)}+2,\dots,n^{(D)},n^{(D)}}\Big)\nonumber\\
&&-\sum_{m=1}^{l-1}\frac{J_m}{2}\sum_{j=\pm1}\Big(a_{\dots,n^{(m)}+j,n^{(m)},\dots,n^{(l)},n^{(l)}+1,\dots}+
a_{\dots,n^{(m)},n^{(m)}+j,\dots,n^{(l)},n^{(l)}+1,\dots}\Big)\nonumber\\
&&-\sum_{m=l+1}^D\frac{J_m}{2}\sum_{j=\pm1}\Big(a_{\dots,n^{(l)},n^{(l)}+1,\dots,n^{(m)}+j,n^{(m)},\dots}+
a_{\dots,n^{(l)},n^{(l)}+1,\dots,n^{(m)},n^{(m)}+j,\dots}\Big)=0,\nonumber\\
&&l=1,\dots,D.
\end{eqnarray}
Here as in (7) and (73)
\begin{equation}
a_{n^{(1)}_2,n^{(1)}_1,\dots n^{(D)}_2,n^{(D)}_1}=a_{n^{(1)}_1,n^{(1)}_2,\dots n^{(D)}_1,n^{(D)}_2}.
\end{equation}

Representing Eq. (145) in the form
\begin{eqnarray}
&&\Big(2\sum_{m=1}^DJ_{z,m}+2\gamma h-E\Big)a_{n^{(1)},n^{(1)},\dots,n^{(l)},n^{(l)}+1,\dots,n^{(D)},n^{(D)}}\nonumber\\
&&-\sum_{m=1}^{l-1}\frac{J_m}{2}\sum_{j=\pm1}\Big(a_{\dots,n^{(m)}+j,n^{(m)},\dots,n^{(l)},n^{(l)}+1,\dots}+
a_{\dots,n^{(m)},n^{(m)}+j,\dots,n^{(l)},n^{(l)}+1,\dots}\Big)\nonumber\\
&&-\frac{J_l}{2}\sum_{j=\pm1}\Big(a_{n^{(1)},n^{(1)},\dots,n^{(l)}+j,n^{(l)}+1,\dots,n^{(D)},n^{(D)}}
+a_{n^{(1)},n^{(1)},\dots,n^{(l)},n^{(l)}+1+j,\dots,n^{(D)},n^{(D)}}\Big)\nonumber\\
&&-\sum_{m=l+1}^D\frac{J_m}{2}\sum_{j=\pm1}\Big(a_{\dots,n^{(l)},n^{(l)}+1,\dots,n^{(m)}+j,n^{(m)},\dots}+
a_{\dots,n^{(l)},n^{(l)}+1,\dots,n^{(m)},n^{(m)}+j,\dots}\Big)\nonumber\\
&&-J_{z,l}a_{n^{(1)},n^{(1)},\dots,n^{(l)},n^{(l)}+1,\dots,n^{(D)},n^{(D)}}+\frac{J_l}{2}\sum_{j=\pm1}\Big(
a_{n^{(1)},n^{(1)},\dots,n^{(l)}+1,n^{(l)}+1,\dots,n^{(D)},n^{(D)}}\nonumber\\
&&+a_{n^{(1)},n^{(1)},\dots,n^{(l)},n^{(l)},\dots,n^{(D)},n^{(D)}}\Big)=0,\qquad l=1,\dots,D,
\end{eqnarray}
we reduce it to Eq. (144) modulo the set of Bethe conditions
\begin{eqnarray}
&&J_l\Big(a_{n^{(1)},n^{(1)},\dots,n^{(D)},n^{(D)}}+a_{n^{(1)},n^{(1)},\dots,n^{(l)}+1,n^{(l)}+1,\dots,n^{(D)},n^{(D)}}\Big)\nonumber\\
&&=2J_{z,l}a_{n^{(1)},n^{(1)},\dots,n^{(l)},n^{(l)}+1,\dots,n^{(D)},n^{(D)}},\qquad l=1,\dots,D,
\end{eqnarray}
which are generalizations of (16) and (79).

Eq. (144) is satisfied for the product wave function
\begin{equation}
a^{(asym,\sigma_1,\dots,\sigma_D)}_{n^{(1)}_1,n^{(1)}_2,\dots,n^{(D)}_1,n^{(D)}_2}(k_{(1)1},k_{(1)2},\dots,k_{(D)1},k_{(D)2})
=\prod_{m=1}^Da^{(\sigma_m)}_{n_1^{(m)},n_2^{(m)}}(k_{(m)1},k_{(m)2}),\qquad
\sigma_m=a,s.
\end{equation}
Eq. (146) for the wave function (149) also will be satisfied if ${\cal N}_a$ defined as the number of $\sigma_m=a$ in (149) is even. Finitely the system of Bethe conditions (148) will be satisfied if ${\cal N}_a\geq2$. Such two-magnon states form the antisymmetric subsector. Their energies are given by the formula
\begin{equation}
E(k_{(1)1},k_{(1)2},\dots,k_{(D)1},k_{(D)2})=\sum_{m=1}^D\Big(E_m(k_{(m)1})+E_m(k_{(m)2})\Big)+2\gamma h,
\end{equation}
where
\begin{equation}
E_m(k)=J_{z,m}-J_m\cos{k}.
\end{equation}

For the $N_1\times\dots\times N_D$ hyper cubic lattice
\begin{equation}
N^{(asym)}={\cal N}(D)\prod_{m=1}^D\frac{N_m(N_m-1)}{2},\qquad N^{(tot)}=\frac{1}{2}\prod_{m=1}^DN_m\Big(\prod_{m=1}^DN_m-1\Big),
\end{equation}
where
\begin{eqnarray}
&&{\cal N}(D)=C_D^2+C_D^4+\dots+C_D^D,\qquad{\rm for}\,\,{\rm even}\,\, D,\\
&&{\cal N}(D)=C_D^2+C_D^4+\dots+C_D^{D-1},\qquad{\rm for}\,\,{\rm odd}\,\, D,
\end{eqnarray}
is the number of all different types of antisymmetric states. In both the cases (153) and (154)
\begin{equation}
{\cal N}(D)=\frac{(1+x)^D+(1-x)^D}{2}\Big|_{x=1}-1=2^{D-1}-1.
\end{equation}
Hence at $N_m\gg1$ ($m=1,\dots,D$) one has
\begin{equation}
N^{(asym)}\approx\Big(1-\frac{1}{2^{D-1}}\Big)N^{(tot)}.
\end{equation}

Wave functions related to the symmetric subsector have a DDD BA form
\begin{eqnarray}
&&a^{(DDD,sym)}_{n^{(1)}_1,n^{(1)}_2,\dots,n^{(D)}_1,n^{(D)}_2}(\{k^{(1)}_{(1)1},k^{(1)}_{(1)2},\dots,k^{(1)}_{(D)1},k^{(1)}_{(D)2}\},\dots,
\{k^{(M)}_{(1)1},k^{(M)}_{(1)2},\dots,k^{(M)}_{(D)1},k^{(M)}_{(D)2}\})\nonumber\\
&&=\sum_{j=1}^MB_ja^{(sym)}_{n^{(1)}_1,n^{(1)}_2,\dots,n^{(D)}_1,n^{(D)}_2}(k^{(j)}_{(1)1},k^{(j)}_{(1)2},\dots,k^{(j)}_{(D)1},k^{(j)}_{(D)2}),
\end{eqnarray}
where
\begin{equation}
a^{(sym)}_{n^{(1)}_1,n^{(1)}_2,\dots,n^{(D)}_1,n^{(D)}_2}(k_{(1)1},k_{(1)2},\dots,k_{(D)1},k_{(D)2})=
\prod_{m=1}^Da^{(s)}_{n_1^{(m)},n_2^{(m)}}(k_{(m)1},k_{(m)2}).
\end{equation}
A substitution of (157) into (148) results in a linear system of equations on the coefficients $B_j$
\begin{equation}
\sum_{j=1}^MB_jX^{(l)}(k_{(l)1}^{(j)},k_{(l)2}^{(j)})=0,\qquad l=1,\dots,D,
\end{equation}
analogous to (57) and (99). In the general case the solution of (159) is
\begin{equation}
B_j=\sum_{l_1,\dots,l_D=1}^{D+1}\varepsilon_{jl_1\dots l_D}\prod_{m=1}^D X^{(m)}(k_{(m)1}^{(l_m)},k_{(m)2}^{(l_m)}),\qquad j=1,\dots,D+1.
\end{equation}
At $D=2,3$ it turns into Eqs. (62) and (100). In the $XX$ case ($J_{z,m}=0$, $m=1,\dots,D$) system (159) again has the solution (60) and hence
\begin{eqnarray}
&&a^{(DDD,sym,XX)}_{n^{(1)}_1,n^{(1)}_2,\dots,n^{(D)}_1,n^{(D)}_2}(k_{(1)1},k_{(1)2},\dots,k_{(D)1},k_{(D)2};
\tilde k_{(1)1},\tilde k_{(1)2},\dots,k_{(D)1},\tilde k_{(D)2})\nonumber\\
&&=\prod_{m=1}^Da^{(s)}_{n_1^{(m)},n_2^{(m)}}(k_{(m)1},k_{(m)2})-\prod_{m=1}^Da^{(s)}_{n_1^{(m)},n_2^{(m)}}(\tilde k_{(m)1},\tilde k_{(m)2}).
\end{eqnarray}
Introducing $2\times2$ matrices
\begin{equation}
K^{(m)}=\left(\begin{array}{cc}
k_{(m)1}&\tilde k_{(m)1}\\
\tilde k_{(m)2}&k_{(m)2}
\end{array}\right),\qquad m=1,\dots D,
\end{equation}
we may represent Eq. (161) in a compact form
\begin{equation}
a^{(DDD,sym,XX)}_{n^{(1)}_1,n^{(1)}_2,\dots,n^{(D)}_1,n^{(D)}_2}(K^{(1)},\dots,K^{(D)})
=\sum_{j,l=1}^2\varepsilon_{jl}\prod_{m=1}^Da^{(s)}_{n_1^{(m)},n_2^{(m)}}(K^{(m)}_{1j},K^{(m)}_{2l}).
\end{equation}
The corresponding energy and components of total quasimomentum (wave number) are given by the formula
\begin{equation}
k^{(m)}=K^{(m)}_{1\sigma(1)}+K^{(m)}_{2\sigma(2)},\quad E=\sum_{m=1}^D\Big(E_m(K^{(m)}_{1\sigma(1)})+E_m(K^{(m)}_{2\sigma(2)})\Big)+2\gamma h,\qquad
\sigma\in{\cal S}_2,
\end{equation}
where ${\cal S}_2$ is the permutation group of two elements.

Though we have not a clear compact proof we suppose that the antisymmetric and symmetric wavefunctions exhaust all the states in the two-magnon sector at arbitrary $D$.

\section{Antisymmetric and symmetric multi-magnon wave functions at arbitrary dimensions}

\subsection{General outlook}

Are the two suggested approaches (related to antisymmetric and symmetric subsectors) powerful only for evaluation of two-magnon $D>1$ states? Can they be extended to multi-magnon (number of magnons $>2$) problems? At the present time we have only some preliminary results on this subject.

First of all (see the next subsection) multi-magnon sectors are not exhausted by antisymmetric and symmetric subsectors. That is why a problem of subdivision of a $Q$-magnon sector on subsectors with respect to symmetries of wave functions is under the question. Nevertheless (see the subsection 7.3) the antisymmetric subsector may be readily obtained within a natural generalization of the approach suggested in the previous sections. It is more difficult to study symmetric states because in each case a necessary amount of different wave vectors is a priori unknown. Nevertheless we have solved this problem for the $XX$ reduction of the $XXZ$ model. An application of the DDD BA even to the three-magnon symmetric subsector of the $XXZ$-model (of course at $D>1$) is also under the question now.

\subsection{${\rm\bf Schr\ddot odinger}$ equation and symmetries of wave functions}

Let us start from a 2D three-magnon state
\begin{equation}
|3\rangle=\sum_{m_1,m_2,m_3,n_1,n_2,n_3}a_{m_1,m_2,m_3,n_1,n_2,n_3}{\bf S}^-_{m_1,n_1}{\bf S}^-_{m_2,n_2}{\bf S}^-_{m_3,n_3}|\emptyset\rangle,
\end{equation}
where
\begin{equation}
a_{m_{\sigma(1)},m_{\sigma(2)},m_{\sigma(3)},n_{\sigma(1)},n_{\sigma(2)},n_{\sigma(3)}}=a_{m_1,m_2,m_3,n_1,n_2,n_3},\qquad\sigma\in{\cal S}_3.
\end{equation}

At $(m_b-m_a)^2+(n_b-n_a)^2>1$, $1\leq a<b\leq3$ the ${\rm Schr\ddot odinger}$ equation for the wave function $a_{m_1,m_2,m_3,n_1,n_2,n_3}$ takes the form
\begin{eqnarray}
&&3(J_{z,1}+J_{z,2}+\gamma h)a_{m_1,m_2,m_3,n_1,n_2,n_3}-\frac{1}{2}\sum_{l=\pm1}\sum_{j=1}^3(J_1a_{\dots,m_j+l,\dots,n_1,n_2,n_3}\nonumber\\
&&+J_2a_{m_1,m_2,m_3,\dots,n_j+l,\dots})
=Ea_{m_1,m_2,m_3,n_1,n_2,n_3}.
\end{eqnarray}
As in the proof of Eq. (148) we may readily show that two-magnon collisions result in the following system of Bethe conditions
\begin{equation}
J_1\Big(a_{m,m,m_3,n,n,n_3}+a_{m+1,m+1,m_3,n,n,n_3}\Big)=2J_{z,1}a_{m,m+1,m_3,n,n,n_3},
\end{equation}
at
\begin{equation}
(m_3-m)^2+(n_3-n)^2>1,\qquad(m_3-m-1)^2+(n_3-n)^2>1,
\end{equation}
(a collision along a horizontal link) and
\begin{equation}
J_2\Big(a_{m,m,m_3,n,n,n_3}+a_{m,m,m_3,n+1,n+1,n_3}\Big)=2J_{z,2}a_{m,m,m_3,n,n+1,n_3},
\end{equation}
at
\begin{equation}
(m_3-m)^2+(n_3-n)^2>1,\qquad(m_3-m)^2+(n_3-n-1)^2>1,
\end{equation}
(a collision along a vertical link), as well as two analogous pairs related to (168) and (170) according to the symmetry (166).

The Bethe conditions related to three-magnon scattering may be reduced to Eqs. (168) and (170) extended to the regions
forbidden by the inequalities (169) and (171). For example the Bethe condition related to three-magnon scattering along the horizontal axis
\begin{eqnarray}
&&J_1\Big(a_{m-1,m-1,m+1,n,n,n}+a_{m,m,m+1,n,n,n}+a_{m-1,m,m,n,n,n}+a_{m-1,m+1,m+1,n,n,n}\Big)\nonumber\\
&&=4J_{z,1}a_{m-1,m,m+1,n,n,n},
\end{eqnarray}
is a linear combination of the two equations
\begin{eqnarray}
J_1\Big(a_{m-1,m-1,m+1,n,n,n}+a_{m,m,m+1,n,n,n}\Big)=2J_{z,1}a_{m-1,m,m+1,n,n,n},\\
J_1\Big(a_{m-1,m,m,n,n,n}+a_{m-1,m+1,m+1,n,n,n}\Big)=2J_{z,1}a_{m-1,m,m+1,n,n,n}.
\end{eqnarray}
Eq. (173) is Eq. (168) shifted on one link to the left. At the same time Eq. (174) may be obtained from Eq. (168) after the permutation: $\sigma(1)=3$, $\sigma(2)=1$, $\sigma(3)=2$. In both the cases inequality (169) is not satisfied. Such reduction of {\it all multi-magnon} Bethe condition to the two-magnon ones (in absence of resonances) is the essence of the BA \cite{1,2,3}.

Contrary to the two-magnon case (Eqs. (16), (79) and (148)) the system of Bethe conditions (168), (170) is not invariant under permutations of one half of wave function indices
\begin{equation}
a_{m_1,m_2,m_3,n_1,n_2,n_3}\longrightarrow a_{m_1,m_2,m_3,n_{\sigma(1)},n_{\sigma(2)},n_{\sigma(3)}},\qquad\sigma\in {\cal S}_3.
\end{equation}
Hence we can not naturally split the three-magnon sector on subsectors related to irreducible representations of the permutation group (for the two-magnon sector these representations correspond to antisymmetric and symmetric wave functions). This obstacle has a clear physical interpretation. Really in the two-magnon case the antisymmetric states correspond to magnons freely moving past each other without scattering. Contrary the symmetric states are related to magnon-magnon collisions. In the three-magnon case we may imagine a situation when two magnons are scattering while the third one moves freely past them. But according to the symmetry (166) we can not exactly identify the scattering pair. For example if we naively put
\begin{equation}
a_{m_1,m_2,m_3,n_1,n_2,n_3}=a_{m_1,m_2,m_3,n_2,n_1,n_3},
\end{equation}
(the particles 1 and 2 are scattering) and
\begin{equation}
a_{m_1,m_2,m_3,n_1,n_2,n_3}=-a_{m_1,m_2,m_3,n_1,n_3,n_2},
\end{equation}
(the particle 3 moves freely past the particle 2) then according to (166) and (177)
\begin{equation}
a_{m_1,m_2,m_3,n_1,n_2,n_3}=a_{m_3,m_2,m_1,n_3,n_2,n_1}=-a_{m_3,m_2,m_1,n_3,n_1,n_2}=-a_{m_1,m_2,m_3,n_2,n_1,n_3},
\end{equation}
in contradiction with (176).

Nevertheless as it will be shown in forward both antisymmetric and symmetric multi-magnon states exist for all $D>1$. Of course they do not
exhaust all the states in the corresponding sectors.

\subsection{Antisymmetric states}

According to the above physical argumentation we may suggest existence of 2D antisymmetric three-magnon states
\begin{equation}
a^{(asym)}_{m_1,m_2,m_3,n_j,n_k,n_l}=\varepsilon_{jkl}a^{(asym)}_{m_1,m_2,m_3,n_1,n_2,n_3},
\end{equation}
related to free moving magnons. Such states really exist and related to the wave functions
\begin{equation}
a^{(asym)}_{m_1,m_2,m_3,n_1,n_2,n_3}(k_1,k_2,k_3,p_1,p_2,p_3)=a^{(a)}_{m_1,m_2,m_3}(k_1,k_2,k_3)a^{(a)}_{n_1,n_2,n_3}(p_1,p_2,p_3),
\end{equation}
where
\begin{equation}
a^{(a)}_{m_1,m_2,m_3}(k_1,k_2,k_3)=\sum_{a,b,c=1}^3\varepsilon_{abc}{\rm e}^{i(k_am_1+k_bm_2+k_cm_3)},
\end{equation}
is a (non normalized) three-magnon wave function of the 1D $XX$ model \cite{15}. Since (181) is antisymmetric under permutations of its wave numbers
\begin{equation}
a^{(a)}_{m_1,m_2,m_3}(k_a,k_b,k_c)=\varepsilon_{abc}a^{(a)}_{m_1,m_2,m_3}(k_1,k_2,k_3),
\end{equation}
we may put as in (34)
\begin{equation}
0\leq k_1<k_2<k_3<2\pi,\qquad0\leq p_1<p_2<p_3<2\pi.
\end{equation}
As in the two-magnon case these antisymmetric states exist also in a finite-periodic $N_x\times N_y$ lattice, where by analogy with Eq. (43) we may put
\begin{eqnarray}
&&k_a=\frac{2\pi j_a}{N_x},\quad a=1,2,3,\qquad0\leq j_1<j_2<j_3<N_x-1,\quad j_1,j_2,j_3\in{\mathbb N},\nonumber\\
&&p_a=\frac{2\pi l_a}{N_y},\quad a=1,2,3,\qquad0\leq l_1<l_2<l_3<N_y-1,\quad l_1,l_2,l_3\in{\mathbb N}.
\end{eqnarray}
The number of these states is
\begin{equation}
N^{(asym)}=\frac{N_xN_y(N_x-1)(N_y-1)(N_x-2)(N_y-2)}{36}.
\end{equation}
At the same time the total number of the three-magnon states is
\begin{equation}
N^{(tot)}=\frac{N_xN_y(N_xN_y-1)(N_xN_y-2)}{6}.
\end{equation}
At $N_x,N_y\gg1$ one has from (185) and (186)
\begin{equation}
N^{(asym)}\approx\frac{N^{(tot)}}{6}.
\end{equation}

Let us now turn to a general $Q$-magnon state in a $D$-dimensional space
\begin{equation}
|Q\rangle=\sum_{n^{(1)}_1,\dots,n^{(1)}_Q;\dots;n^{(D)}_1,\dots,n^{(D)}_Q}a_{n^{(1)}_1,\dots,n^{(1)}_Q;\dots;n^{(D)}_1,\dots,n^{(D)}_Q}
\prod_{l=1}^Q{\bf S}^-_{n^{(1)}_l,\dots,n^{(D)}_l}|\emptyset\rangle,
\end{equation}
where as in (166)
\begin{equation}
a_{n^{(1)}_{\sigma(1)},\dots,n^{(1)}_{\sigma(Q)};\dots;n^{(D)}_{\sigma(1)},\dots,n^{(D)}_{\sigma(Q)}}=
a_{n^{(1)}_1,\dots,n^{(1)}_Q;\dots;n^{(D)}_1,\dots,n^{(D)}_Q},\qquad\sigma\in{\cal S}_Q.
\end{equation}

Following the same argumentation as was given in evaluation of Eqs. (144) and (148) we may show that the ${\rm Schr\ddot odinger}$ equation for the wave function splits on the system
\begin{eqnarray}
&&\Big(Q\sum_{m=1}^DJ_{z,m}+Q\gamma h-E\Big)a_{n^{(1)}_1,\dots,n^{(1)}_Q;\dots;n^{(D)}_1,\dots,n^{(D)}_Q}
-\sum_{m=1}^D\frac{J_m}{2}\sum_{l=1}^Q\sum_{j=\pm1}a_{\dots,n^{(m)}_l+j,\dots}=0,\nonumber\\
&&
\sum_{m=1}^D\Big(n^{(m)}_l-n^{(m)}_j\Big)^2>1,\qquad1\leq j<l\leq Q,
\end{eqnarray}
related to free motion and the set of Bethe conditions related to pair scattering. According the symmetry (189) we may limit ourselves by scattering of the particles with numbers 1 and 2, when a collision along the $l$-th Cartesian axis gives
\begin{eqnarray}
&&J_l\Big(a_{n^{(1)},n^{(1)}\dots;\dots;n^{(l)},n^{(l)}\dots;\dots;n^{(D)},n^{(D)},\dots}+
a_{n^{(1)},n^{(1)},\dots;\dots;n^{(l)}+1,n^{(l)}+1,\dots;\dots;n^{(D)},n^{(D),\dots}}\Big)
\nonumber\\
&&=2J_{z,l}a_{n^{(1)},n^{(1)},\dots;\dots;n^{(l)},n^{(l)}+1,\dots;\dots;n^{(D)},n^{(D),\dots}},\qquad l=1,\dots,D.
\end{eqnarray}

We suggest now the following set of $D$-dimensional $Q$-magnon antisymmetric wave functions
\begin{eqnarray}
&&a^{(asym,\sigma_1,\dots,\sigma_D)}_{n^{(1)}_1,\dots,n^{(1)}_Q;\dots;n^{(D)}_1,\dots,n^{(D)}_Q}(k_{(1)1},\dots k_{(1)Q};\dots;k_{(D)1},\dots k_{(D)Q})
\nonumber\\
&&=\prod_{j=1}^D
a^{(\sigma_j)}_{n^{(j)}_1,\dots,n^{(j)}_Q}(k_{(j)1},\dots k_{(j)Q}),\qquad
\sigma_j=a,s,
\end{eqnarray}
where
\begin{equation}
a^{(a)}_{n_1,\dots,n_Q}(k_1,\dots k_Q)=\sum_{a_1,\dots,a_Q=1}^Q\varepsilon_{a_1\dots a_Q}{\rm e}^{i\sum_{j=1}^Qk_{a_j}n_j},
\end{equation}
is a (non normalized) $Q$-magnon wave function of the 1D $XX$ model \cite{15} and
\begin{equation}
a^{(s)}_{n_1,\dots,n_Q}(k_1,\dots k_Q)=\sum_{\sigma\in{\cal S}_Q}{\rm e}^{i\sum_{j=1}^Qk_{\sigma(j)}n_j}.
\end{equation}
As in Eq. (149) the parameter ${\cal N}_a$ (the number of $\sigma_j=a$ in (192)) should be even in order to ensure the symmetry condition (189)
and strictly positive to ensure the property
\begin{equation}
a^{(asym,\sigma_1,\dots,\sigma_D)}_{n^{(1)},n^{(1)},\dots;\dots;n^{(l)}_1,n^{(l)}_2,\dots;\dots;n^{(D)},n^{(D),\dots}}=0,
\end{equation}
(in Eq. (195) $n^{(j)}_1=n^{(j)}_2$ for all $1\leq j\leq D$ except may be $j=l$) directly resulting in Eq. (191).

Following the previous argumentation (see Eqs. (152)-(155)) one may readily prove that in this case
\begin{equation}
N^{(asym)}(Q,D)=(2^{D-1}-1)\prod_{j=1}^DC_{N_j}^Q,\qquad N^{(tot)}(Q,D)=C_{\prod_{j=1}^DN_j}^Q,
\end{equation}
and at $N_j\gg1$, $j=1,\dots,D$
\begin{equation}
N^{(asym)}(Q,D)\approx\frac{2^{D-1}-1}{(Q!)^D}\Big(\prod_{j=1}^DN_j\Big)^Q,\qquad N^{(tot)}(Q,D)\approx\frac{1}{Q!}\Big(\prod_{j=1}^DN_j\Big)^Q.
\end{equation}
Hence at $N_j\gg1$, $j=1,\dots,D$
\begin{equation}
N^{(asym)}(Q,D)\approx\frac{2^{D-1}-1}{(Q!)^{D-1}}N^{(tot)}(Q,D).
\end{equation}

\subsection{Symmetric states in the XX case}

In the XX ($J_{z,m}$, $m=1,\dots D$) case natural generalizations of $D=2,3$ two-magnon symmetric wave functions for arbitrary $D$ and $Q$ may be readily obtained within the DDD BA machinery.
Starting from $D=2$ and $Q=3$ we suggest the following symmetric three-magnon wave function
\begin{eqnarray}
&&a^{(DDD,sym,XX)}_{m_1,m_2,m_3,n_1,n_2,n_3}(k_1^{(1)},k_2^{(1)},k_3^{(1)},p_1^{(1)},p_2^{(1)},p_3^{(1)};\dots;
k_1^{(3)},k_2^{(3)},k_3^{(3)},p_1^{(3)},p_2^{(3)},p_3^{(3)})\nonumber\\
&&=\sum_{a,b,c=1}^3\varepsilon_{abc}
a^{(sym)}_{m_1,m_2,m_3,n_1,n_2,n_3}(k_1^{(a)},k_2^{(b)},k_3^{(c)},p_1^{(a)},p_2^{(b)},p_3^{(c)}),
\end{eqnarray}
where
\begin{equation}
a^{(sym)}_{m_1,m_2,m_3,n_1,n_2,n_3}(k_1,k_2,k_3,p_1,p_2,p_3)=a^{(s)}_{m_1,m_2,m_3}(k_1,k_2,k_3)a^{(s)}_{n_1,n_2,n_3}(p_1,p_2,p_3),
\end{equation}
and according to energy and quasimomentum conservations laws
\begin{eqnarray}
&&k_1^{(a)}+k_2^{(b)}+k_3^{(c)}=k,\qquad p_1^{(a)}+p_2^{(b)}+p_3^{(c)}=p,\nonumber\\
&&E_1(k_1^{(a)})+E_1(k_2^{(b)})+E_1(k_3^{(c)})+E_2(p_1^{(a)})+E_2(p_2^{(b)})+E_2(p_3^{(c)})+3\gamma h=E.
\end{eqnarray}
Here $k$, $p$ and $E$ are components of the total quasimomentum and energy. In all the sums of Eq. (201) there should be $a\neq b\neq c\neq a$.

Introducing the $3\times3$ matrices
\begin{equation}
K=\left(\begin{array}{ccc}
k_1^{(1)}&k_1^{(2)}&k_1^{(3)}\\
k_2^{(1)}&k_2^{(2)}&k_2^{(3)}\\
k_3^{(1)}&k_3^{(2)}&k_3^{(3)}
\end{array}\right),\qquad
P=\left(\begin{array}{ccc}
p_1^{(1)}&p_1^{(2)}&p_1^{(3)}\\
p_2^{(1)}&p_2^{(2)}&p_2^{(3)}\\
p_3^{(1)}&p_3^{(2)}&p_3^{(3)}
\end{array}\right),
\end{equation}
we may reduce Eq. (199) to the form similar to (163)
\begin{eqnarray}
a^{(DDD,sym,XX)}_{m_1,m_2,m_3,n_1,n_2,n_3}(K,P)=\sum_{a,b,c=1}^3\varepsilon_{abc}a^{(s)}_{m_1,m_2,m_3}(K_{1a},K_{2b},K_{3c})
a^{(s)}_{n_1,n_2,n_3}(P_{1a},P_{2b},P_{3c})
\end{eqnarray}
The energy-quasimomentum relations (201) will take the form similar to (164)
\begin{equation}
k=\sum_{j=1}^3K_{j\sigma(j)},\quad p=\sum_{j=1}^3P_{j\sigma(j)},\quad E=\sum_{j=1}^3\Big(E_1(K_{j\sigma(j)})+E_2(P_{j\sigma(j)})\Big)+3\gamma h,\qquad
\sigma\in{\cal S}_3.
\end{equation}

Since
\begin{eqnarray}
&&a^{(DDD,sym,XX)}_{m,m,m_3,n,n,n_3}(K,P)=\frac{1}{4}\sum_{a,b,c=1}^3\varepsilon_{abc}\sum_{\sigma,\tilde\sigma\in{\cal S}_3}\Big({\rm e}^{i(K_{\sigma(1)a}+K_{\sigma(2)b})m}+{\rm e}^{i(K_{\sigma(2)a}+K_{\sigma(1)b})m}\Big)\nonumber\\
&&\cdot\Big({\rm e}^{i(P_{\tilde\sigma(1)a}+P_{\tilde\sigma(2)b})n}+{\rm e}^{i(P_{\tilde\sigma(2)a}+P_{\tilde\sigma(1)b})n}\Big)
{\rm e}^{iK_{\sigma(3)c}m_3}{\rm e}^{iP_{\tilde\sigma(3)c}n_3}=0,
\end{eqnarray}
the wave function (199) solves also the Bethe condition (173). In the same manner it may be proved that the condition (174) also is satisfied.

Turning to the general $D$-dimensional $Q$-magnon problem we change the $3\times3$ matrices $K$ and $P$ on the set of $Q\times Q$ matrices $K^{(j)}$,
$j=1,\dots,D$ and suggest the general formula
\begin{equation}
a^{(DDD,sym,XX)}_{n^{(1)}_1,\dots,n^{(1)}_Q;\dots;n^{(D)}_1,\dots,n^{(D)}_Q}(K^{(1)},\dots,K^{(D)})=\sum_{j_1,\dots,j_Q=1}^Q
\varepsilon_{j_1\dots j_Q}\prod_{m=1}^D
a^{(s)}_{n^{(m)}_1,\dots,n^{(m)}_Q}(K^{(m)}_{1j_1},\dots,K^{(m)}_{Qj_Q}),
\end{equation}
which turns into (203) at $D=2$, $Q=3$ and in fact is equivalent to (61) and (102) at $D=2$ and $D=3$ ($Q=2$).

As it was previously done in the case $D=2$, $Q=3$ we may readily prove analogously to (205) that
\begin{equation}
a^{(DDD,sym,XX)}_{n^{(1)},n^{(1)},n^{(1)}_3,\dots,n^{(1)}_Q;\dots;n^{(D)},n^{(D)},n^{(D)}_3,\dots,n^{(D)}_Q}=0,
\end{equation}
(in Eq. (207) $n^{(j)}_1=n^{(j)}_2$ for all $1\leq j\leq D$) which in the $XX$ case yields (191).

\section{The two-magnon problem in the Dyson-Maleev and Holstein-Primakoff approaches}

Since the 2D and 3D ferromagnets has been previously studied by a numerous of alternative approaches we need to compare the two most popular of them with the DDD BA. For simplicity we shall limit ourselves by the 2D case.

In his pioneering paper \cite{9} F. J. Dyson utilized the two-magnon wave functions
\begin{equation}
a^{(D)}_{m_1,m_2,n_1,n_2}(k_1,k_2,p_1,p_2)={\rm e}^{i(k_1m_1+k_2m_2+p_1n_1+p_2n_2)}+{\rm e}^{i(k_2m_1+k_1m_2+p_2n_1+p_1n_2)}.
\end{equation}
The corresponding states are neither eigenvectors of the Hamiltonian, no orthogonal to each other. The former was interpreted as a manifestation of the dynamical while the latter of the kinematical interactions between magnons.
In the present paper we have used the following linear combinations of the Dyson's wave functions (see Eqs. (17), (21) and (50), (51))
\begin{eqnarray}
a^{(asym)}_{m_1,m_2,n_1,n_2}(k_1,k_2,p_1,p_2)=a^{(D)}_{m_1,m_2,n_1,n_2}(k_1,k_2,p_1,p_2)-a^{(D)}_{m_1,m_2,n_1,n_2}(k_2,k_1,p_1,p_2),\\
a^{(sym)}_{m_1,m_2,n_1,n_2}(k_1,k_2,p_1,p_2)=a^{(D)}_{m_1,m_2,n_1,n_2}(k_1,k_2,p_1,p_2)+a^{(D)}_{m_1,m_2,n_1,n_2}(k_2,k_1,p_1,p_2).
\end{eqnarray}

The wave functions (209) are related to the antisymmetric subsector for which both the dynamical and kinematical interactions cancel. In the symmetric subsector the dynamical interaction cancels for DDD BA states (54), (62) and (61) however its kinematical counterpart becomes more complicated and resulting for example in the appearance of the bound and resonant states \cite{10,11,12} (which according to (39) are away from the antisymmetric sector).

F. J. Dyson also suggested for the ferromagnetic spin system an effective bosonic Hamiltonian which acts in an extended Hilbert space. Let us show that this Hamiltonian correctly reproduce the antisymmetric two-magnon subsector.

The most direct transition from (1) to the Dyson's effective Hamiltonian follows from the Maleev representation for spin operators \cite{13}
\begin{equation}
{\bf\tilde S}^+=(1-c^{\dagger}c)c,\qquad{\bf\tilde S}^z=\frac{1}{2}-c^{\dagger}c,\qquad{\bf\tilde S}^-=c^{\dagger},
\end{equation}
where $c$ and $c^{\dagger}$ is a pair of Bose operators
\begin{equation}
[c,c^{\dagger}]=1.
\end{equation}
The representation (211) reproduces only a part of algebraic relations between the spin operators. Namely, it may be readily checked that
\begin{equation}
[{\bf\tilde S}^+,{\bf\tilde S}^-]=2{\bf\tilde S}^z,\qquad[{\bf\tilde S}^z,{\bf\tilde S}^{\pm}]=\pm{\bf\tilde S}^{\pm},
\qquad \frac{1}{2}\Big({\bf\tilde S}^+{\bf\tilde S}^-+{\bf\tilde S}^-{\bf\tilde S}^+\Big)+({\bf\tilde S}^z)^2=
\frac{3}{4},
\end{equation}
however
\begin{equation}
({\bf\tilde S}^-)^2\neq0,
\end{equation}
(and also $({\bf\tilde S}^+)^2\neq0$, $({\bf\tilde S}^-)^{\dagger}\neq{\bf\tilde S}^+$).

Let $\hat H^{(D-M)}$ be a 2D bosonic Dyson-Maleev Hamiltonian obtained from (1) under the substitution
\begin{equation}
{\bf S}_{m,n}\longrightarrow{\bf\tilde S}_{m,n}.
\end{equation}
The ground state (2) turns into the state $|\emptyset\rangle^{(D-M)}$ defined by the following conditions
\begin{equation}
c_{m,n}|\emptyset\rangle^{(D-M)}=0\Longrightarrow{\bf\tilde S}_{m,n}^+|\emptyset\rangle^{(D-M)}=0,\qquad m,n=-\infty\dots\infty.
\end{equation}

The bosonic Dyson-Maleev Hamiltonian $\hat H^{(D-M)}$ has the form \cite{9,13}
\begin{equation}
\hat H^{(D-M)}=\hat H_0^{(D-M)}+\hat V^{(D-M)},
\end{equation}
where
\begin{eqnarray}
&&\hat H_0^{(D-M)}=\sum_{m,n}\Big((J_{z,1}+J_{z,2}+\gamma h)c^{\dagger}_{m,n}
-\frac{1}{2}\sum_{\sigma=\pm1}(J_1c^{\dagger}_{m+\sigma,n}
+J_2c^{\dagger}_{m,n+\sigma})\Big)c_{m,n},\\
&&\hat V^{(D-M)}=\sum_{m,n}\Big(\frac{1}{2}\sum_{\sigma=\pm1}c^{\dagger}_{m,n}(J_1c^{\dagger}_{m+\sigma,n}+
J_2c^{\dagger}_{m,n+\sigma})c^2_{m,n}
\nonumber\\
&&-c^{\dagger}_{m,n}c_{m,n}(J_{z,1}c^{\dagger}_{m+1,n}c_{m+1,n}+J_{z,2}c^{\dagger}_{m,n+1}c_{m,n+1})\Big).
\end{eqnarray}

A one-magnon state for $\hat H^{(D-M)}$ is similar to (3)
\begin{equation}
|k,p\rangle^{(D-M)}=\sum_{m,n}{\rm e}^{i(km+pn)}{\bf\tilde S}^-_{m,n}|\emptyset\rangle^{(D-M)}=
\sum_{m,n}{\rm e}^{i(km+pn)}c^{\dagger}_{m,n}|\emptyset\rangle^{(D-M)}
\end{equation}
and its energy is given by Eq. (4). Hence the one-magnon spectrums of $\hat H$ and $\hat H^{(D-M)}$ coincide.

A two-magnon state should be
\begin{equation}
|2\rangle^{(D-M)}=\sum_{m_1,m_2,n_1,n_2}a_{m_1,m_2,n_1,n_2}^{(D-M)}c^{\dagger}_{m_1,n_1}c^{\dagger}_{m_2,n_2}
|\emptyset\rangle^{(D-M)},
\end{equation}
where without lost of generality the wave function $a^{(D-M)}_{m_1,m_2,n_1,n_2}$ satisfies the symmetry condition
\begin{equation}
a^{(D-M)}_{m_2,m_1,n_2,n_1}=a^{(D-M)}_{m_1,m_2,n_1,n_2},
\end{equation}
analogous to (7).

Taking into account that
\begin{equation}
c^2_{m,n}|2\rangle^{(D-M)}=2a_{m,m,n,n}^{(D-M)}|\emptyset\rangle^{(D-M)},
\end{equation}
and that the term proportional to $c^{\dagger}_{m,n}c^{\dagger}_{m+\sigma,n}|\emptyset\rangle$ appears in (221) twice (as $c^{\dagger}_{m,n}c^{\dagger}_{m+\sigma,n}|\emptyset\rangle$ and as $c^{\dagger}_{m+\sigma,n}c^{\dagger}_{m,n}|\emptyset\rangle$) and analogous is true for $c^{\dagger}_{m,n}c^{\dagger}_{m,n+\sigma}|\emptyset\rangle$,
one gets the following ${\rm Schr\ddot odinger}$ equation
\begin{eqnarray}
&&\Big((2-\delta_{|m_2-m_1|,1}\delta_{n_1n_2})J_{z,1}+(2-\delta_{m_1m_2}\delta_{|n_2-n_1|,1})J_{z,2}+2\gamma h\Big)a_{m_1,m_2,n_1,n_2}^{(D-M)}\nonumber\\
&&-\frac{1}{2}
\sum_{\sigma=\pm1}\Big(J_1(a_{m_1+\sigma,m_2,n_1,n_2}^{(D-M)}
+a_{m_1,m_2+\sigma,n_1,n_2}^{(D-M)})+J_2(a_{m_1,m_2,n_1+\sigma,n_2}^{(D-M)}+a_{m_1,m_2,n_1,n_2+\sigma}^{(D-M)})\Big)\nonumber\\
&&+\frac{1}{2}\Big(\delta_{|m_2-m_1|,1}\delta_{n_1n_2}J_1
+\delta_{m_1m_2}\delta_{|n_2-n_1|,1}J_2\Big)\Big(a^{(D-M)}_{m_1,m_1,n_1,n_1}+a^{(D-M)}_{m_2,m_2,n_2,n_2}\Big)\nonumber\\
&&=Ea_{m_1,m_2,n_1,n_2}^{(D-M)}.
\end{eqnarray}

Like the system (13)-(15) Eq. (224) is invariant under the substitution
\begin{equation}
a_{m_1,m_2,n_1,n_2}^{(D-M)}\longrightarrow a_{m_2,m_1,n_1,n_2}^{(D-M)},\qquad
a_{m_1,m_2,n_1,n_2}^{(D-M)}\longrightarrow a_{m_1,m_2,n_2,n_1}^{(D-M)},
\end{equation}
and hence also supports a subdivision of the two-magnon sector on the antisymmetric and symmetric subsectors. Moreover it may be readily seen that for the antisymmetric subsector both Eq. (224) and the system (13)-(15) are equivalent.
However in the symmetric sector Eq. (224) contrary to Eqs. (13)-(15) mixes the spurious components $a^{(D-M)}_{m,m,n,n}$ with the physical ones.
Hence the Dyson-Maleev effective Hamiltonian correctly reproduce only the results related to the antisymmetric two-magnon subsector.

The subdivision on antisymmetric and symmetric subsectors is also valid within the Holstein-Primakoff approach \cite{14}, based on the following representation of spin-S operators
\begin{equation}
{\bf S}^+=\sqrt{2S-c^{\dagger}c}\cdot c,\qquad{\bf S}^-=c^{\dagger}\sqrt{2S-c^{\dagger}c},\qquad{\bf S}^z=S-c^{\dagger}c,
\end{equation}
($c$ and $c^{\dagger}$ are the same as in Eq. (212)). The latter is more effective at $S\geq1$ when one may use the series expansion
\begin{equation}
\sqrt{2S-c^{\dagger}c}=\sqrt{2S}\Big(1-\frac{c^{\dagger}c}{4S}+\dots\Big).
\end{equation}
For $S=1/2$ ferromagnet at {\it low density of magnons} (at low temperatures) in order to get out from the radicals Holstein and Primakoff suggested to put
\begin{equation}
{\bf S}^+\approx c,\qquad{\bf S}^-\approx c^{\dagger},\qquad{\bf S}^z=\frac{1}{2}-c^{\dagger}c.
\end{equation}
The main advantage of such rude approximation is its universal and rather satisfactory applicability to a lot of models more complex than the one considered here (for example including dipole-dipole long-range interactions or magnetic field oriented not along the symmetry axis). For our model a substitution of (228) into (1) results in the Hamiltonian
\begin{equation}
\hat H^{(H-P)}=\hat H_0^{(H-P)}+\hat V^{(H-P)},
\end{equation}
where
\begin{eqnarray}
&&\hat H_0^{(H-P)}=\sum_{m,n=-\infty}^{\infty}\Big[-\frac{J_1}{2}\Big(c^{\dagger}_{m-1,n}+c^{\dagger}_{m+1,n}\Big)c_{m,n}-
\frac{J_2}{2}\Big(c^{\dagger}_{m,n-1}+c^{\dagger}_{m,n+1}\Big)c_{m,n}\nonumber\\
&&+(J_{z,1}+J_{z,2}+\gamma h)c^{\dagger}_{m,n}c_{m,n}\Big],\\
&&\hat V^{(H-P)}=-\sum_{m,n=-\infty}^{\infty}c^{\dagger}_{m,n}c_{m,n}(J_{z,1}c^{\dagger}_{m+1,n}c_{m+1,n}+J_{z,2}c^{\dagger}_{m,n+1}c_{m,n+1}).
\end{eqnarray}
The Holstein-Primakoff ground state $|\emptyset\rangle^{(H-P)}$ and one-magnon states coincide with the corresponding Dyson-Maleev ones (216) and (220).

For a two-magnon state
\begin{equation}
|2\rangle^{(H-P)}=\sum_{m_1,m_2,n_1,n_2}a_{m_1,m_2,n_1,n_2}^{(H-P)}c^{\dagger}_{m_1,n_1}c^{\dagger}_{m_2,n_2}
|\emptyset\rangle^{(H-P)},
\end{equation}
(where we again suggest the general symmetry condition (7)) the ${\rm Schr\ddot odinger}$ equation takes the form
\begin{eqnarray}
&&\Big((2-\delta_{|m_2-m_1|,1}\delta_{n_1n_2})J_{z,1}+(2-\delta_{m_1m_2}\delta_{|n_2-n_1|,1})J_{z,2}+2\gamma h\Big)a_{m_1,m_2,n_1,n_2}^{(H-P)}\nonumber\\
&&-\frac{1}{2}
\sum_{\sigma=\pm1}\Big(J_1(a_{m_1+\sigma,m_2,n_1,n_2}^{(H-P)}
+a_{m_1,m_2+\sigma,n_1,n_2}^{(H-P)})+J_2(a_{m_1,m_2,n_1+\sigma,n_2}^{(H-P)}+a_{m_1,m_2,n_1,n_2+\sigma}^{(H-P)})\Big)\nonumber\\
&&=Ea_{m_1,m_2,n_1,n_2}^{(H-P)}.
\end{eqnarray}
At $(m_2-m_1)^2+(n_2-n_1)^2>1$ Eq. (233) is equivalent to Eq. (13) but for $m_2=m_1$, $n_2=n_1\pm1$ and $m_2=m_1\pm1$, $n_2=n_1$ it also contains the spurious terms proportional to $a_{mm,nn}$. In the same manner as it was done in the Dyson-Maleev case we may readily prove that a subdivision on the antisymmetric and symmetric subsectors is also valid for Eq. (233) which however correctly reproduces only the antisymmetric subsector.

\section{Summary and discussion}

In the present paper we have used the Bethe ansatz paradigm for evaluation of multi-magnon scattering states in ($D>1$)-dimensional Heisenberg-Ising ferromagnet and its $XX$ reduction. In the two-magnon case we have shown that the corresponding Fock space is subdivided on two orthogonal subsectors.

The antisymmetric subsector corresponds to pairs of non-interacting magnons and may be completely described by the direct $D$-dimensional generalization of the traditional BA for both finite-periodic and infinite lattices. This subsector contains about 1/2 of all two-magnon states for a big 2D square lattice and about 3/4 for a big 3D simple cubic lattice (for general antisymmetric $Q$-magnon states on a $D$-dimensional hyper cubic lattice see Eq (198)). The corresponding wave functions have universal 2D or 3D generalized Bethe forms (see Eqs. (22), (85) and (192) for general $Q$ and $D$) which does not depend on the anisotropy parameters. Unlike the 1D case where a two-magnon wave function is a sum of two exponents, the obtained wave functions contain four ones in the 2D case and eight in the 3D (from Eqs. (192) and (193) follows that at general $Q$ and $D$ the number of exponents is $(Q!)^D$).

The symmetric subsector related to interacting magnons is much more complicated and just absorbs all the difficulties of the problem. For infinite lattices (where the periodicity conditions (40) do not result in the quantization rules (43)) it may be effectively studied within the DDD BA. Namely in the general case $(J_{z,1},J_{z,2},J_{z,3}\neq0)$ we have obtained the overcomplete systems of DDD BA two-magnon states with the DDD BA parameter $M=3$  in 2D (54), (62) and $M=4$ in 3D (97), (100) (really $M=D+1$ for an arbitrary $D>1$ (157), (160)). The reduced XX case (58), (101) may be treated within the $M=2$ DDD BA at all $D$ (61), (102).

In order to clarify perspectives of the two suggested approaches we briefly studied antisymmetric and symmetric subsectors of $(Q>2)$-magnon sectors (in this cases the total $Q$-magnon sector is not exhausted by these two subsectors). In this way we have obtained the full description of the antisymmetric subsectors. However for the symmetric one we have succeeded only for the $XX$ reduction of the model (see Eq. (206)). The obtained classes of multi-magnon states may be interpreted as towers of exactly solvable states for non-integrable ($XXZ$ or $XX$) model in the sense of Ref. 21 (where the analogous tower was obtained for the AKLT chain \cite{22}).

Since the 2D and 3D Heisenberg-Ising ferromagnets (1) and (68) has been already successfully studied within a various number of alternative approximative approaches
\cite{9,13,14} we need to emphasis corresponding impacts of the present paper. The mathematical impact is the suggestion of the two generalized versions of the usual 1D Bethe ansatz. The first one effective for integrable systems was applied to antisymmetric subsectors. The second one effective for {\it non-integrable} systems was applied to symmetric subsectors. This is the main result of the paper which continues the line of research started in \cite{8}. The physical impact is the subdivision of the two-magnon sector on two subsectors related to moving pass each other (the antisymmetric subsector) and colliding (the symmetric one) magnons. Analyzing the {\it exact} wave functions we have concluded that the antisymmetric subsector corresponds to regular, while symmetric to chaotic quantum dynamics. As an accessory result we have shown that the difference between exact and approximative Dyson-Maleev and Holstein-Primakoff approaches lies just in the treatment of the symmetric subsector.

Finitely we again notice that for evaluation of low temperature thermodynamical quantities \cite{23,24} it is necessary to obtain the resolution of unity for the two-magnon sector in the similar manner that it was done in the 1D case \cite{25}. The corresponding resolutions for the 2D and 3D antisymmetric sectors are given by Eqs. (39) and (94). However for the symmetric sectors they are unknown.

\end{document}